\documentclass[journal]{IEEEtran}
\usepackage{amsmath,amsfonts}
\usepackage{algorithmic}
\usepackage{fontawesome5}

\usepackage{array}
\usepackage[caption=false,font=normalsize,labelfont=sf,textfont=sf]{subfig}
\usepackage{textcomp}
\usepackage{stfloats}
\usepackage{hhline}
\usepackage{url}
\usepackage{verbatim}
\usepackage{graphicx}
\def\BibTeX{{\rm B\kern-.05em{\sc i\kern-.025em b}\kern-.08em
    T\kern-.1667em\lower.7ex\hbox{E}\kern-.125emX}}
\usepackage{balance}

\usepackage[table,dvipsnames]{xcolor}
\usepackage[shrink=25]{microtype}
\usepackage{tikz}
\usepackage{pgfplots}
\pgfplotsset{compat=1.18}
\usetikzlibrary{patterns}
\usetikzlibrary[shadows]
\usepackage{ifthen}
\usetikzlibrary{calc}
\usetikzlibrary{arrows.meta}
\usetikzlibrary{external}
\usetikzlibrary{positioning,fit,calc}
\usepackage[mathscr]{eucal}
\usepackage{dsfont}
\usepackage{scalefnt}
\usepackage{amsmath,amssymb,amsfonts}
\usepackage{algorithmic}
\usepackage{graphicx}
\usepackage{textcomp}
\usepackage{listings}
\definecolor{backcolour}{rgb}{0.95,0.95,0.92}
\definecolor{arrowcolor}{RGB}{145,148,138}
\definecolor{weborange}{RGB}{255,165,0}
\definecolor{darkblue}{rgb}{0.0,0.0,0.6}
\definecolor{cyan}{rgb}{0.0,0.6,0.6}
\definecolor{deepred}{rgb}{0.6,0,0}
\definecolor{deepgreen}{rgb}{0,0.5,0}

\lstdefinelanguage{PAULA}{
  keywords={par, and, or, if, ifrt, in, out, program, variable, parameter},
  emph={matrix_vector_multiplication, mvm},
  emphstyle=\color{deepred},
  comment=[l]{//},
  commentstyle=\color{green!50!black},
  keywordstyle=\color{blue}
}

\lstdefinestyle{my_PAULA_style}{
  language=PAULA,
  backgroundcolor=\color{backcolour},
  breaklines=true,
  xleftmargin=1.5em,
  numbers=left,
  numbersep=0.5em
}

\lstdefinestyle{PAULA_style}{
  language=PAULA,
  breaklines=true
}

\lstdefinestyle{PseudoCode}{%
  keywords={for, do},
  breaklines=true,
  backgroundcolor=\color{backcolour},
  keywordstyle=\color{blue}
}

\lstdefinestyle{log}{%
  breaklines=true,
  backgroundcolor=\color{backcolour}
}

\lstdefinelanguage{json}{%
  basicstyle=\normalfont\ttfamily,
  breaklines=true,
  backgroundcolor=\color{backcolour},
  literate=
     *{0}{{{\color{deepred}0}}}{1}
      {1}{{{\color{deepred}1}}}{1}
      {2}{{{\color{deepred}2}}}{1}
      {3}{{{\color{deepred}3}}}{1}
      {4}{{{\color{deepred}4}}}{1}
      {5}{{{\color{deepred}5}}}{1}
      {6}{{{\color{deepred}6}}}{1}
      {7}{{{\color{deepred}7}}}{1}
      {8}{{{\color{deepred}8}}}{1}
      {9}{{{\color{deepred}9}}}{1}
      {:}{{{\color{deepred}{:}}}}{1}
      {,}{{{\color{deepred}{,}}}}{1}
      {\{}{{{\color{darkblue}{\{}}}}{1}
      {\}}{{{\color{darkblue}{\}}}}}{1}
      {[}{{{\color{darkblue}{[}}}}{1}
      {]}{{{\color{darkblue}{]}}}}{1},
}

\lstdefinestyle{myC}{%
  language=C,
  tabsize=2,
  keywordstyle=\color{blue},
  commentstyle=\color{green!50!black},
  stringstyle=\color{deepgreen},
  backgroundcolor=\color{backcolour},
  breaklines=true,
  breakatwhitespace=true,
  moredelim=[s][\color{red}]{\$\{}{\}},
  postbreak=\mbox{\textcolor{arrowcolor}{$\hookrightarrow$}}
}
\lstdefinestyle{myC2}{
  style=myC,
  keywords=[2]{mapping},
  keywordstyle=[2]\color{cyan},
  emphstyle=\color{deepred}
}

\usepackage{multirow}
\usepackage{pgfplotstable}
\usepackage{pgfplots}
\usepackage{marvosym}
\usepackage[english]{babel}
\usepackage{upgreek}
\usepackage{makecell}
\usepackage{framed}
\usepackage{fp}
\usepackage{setspace}
\usepackage{pifont}
\usepackage[nolist]{acronym}

\def\BibTeX{{\rm B\kern-.05em{\sc i\kern-.025em b}\kern-.08em
    T\kern-.1667em\lower.7ex\hbox{E}\kern-.125emX}}

\usepackage[hidelinks]{hyperref}
\usepackage[backend=biber, style=numeric, hyperref, maxnames=6, minnames=1, giveninits=true, maxcitenames=2, mincitenames=1,bibencoding=utf8, isbn=true,sorting=none]{biblatex}
\usepackage{cleveref}
\usepackage{siunitx}
\usepackage{xspace}
\usepackage{xargs}

\usepackage{csquotes}

\newcommand{\II}{\textit{II}\xspace}

\DeclareSIUnit\flops{FLOPS}
\DeclareSIUnit\flop{FLOP}

\DeclareSIUnit\ops{OPS}
\DeclareSIUnit\op{OP}

\pgfplotsset{compat=1.18}

\graphicspath{{figures/}}

\usepackage{bm}

\newcommand{\mmat}[1]{\mathbf{#1}}
\newcommand{\mvec}[1]{\bm{#1}}
\newcommand{\mset}[1]{\mathcal{#1}}
\newcommand{\mtrans}{^\intercal}
\newcommand{\mbaseSet}[1]{\mathbb{#1}}
\newcommand{\msetBuilder}[2]{\{{#1}\ |\ {#2}\}}

\newcommand{\plaDimension}{n}

\newcommand{\plaGenericVector}[1]{(#1_0, \dots, #1_{\plaDimension - 1})}

\newcommand{\plaDimensionVector}[2]{#1_0 \times \dots \times #1_{#2} \times \dots \times #1_{\plaDimension - 1}}

\newcommand{\plaIteration}{\mvec{i}}
\newcommand{\plaIntraIteration}{\mvec{j}}
\newcommand{\plaInterIteration}{\mvec{k}}

\newcommand{\plaIterationSpace}{\mset{I}}
\newcommand{\plaIterationSpaceDefinition}{\plaIterationSpace \subseteq \mbaseSet{Z}^\plaDimension}
\newcommand{\plaIterationLongDefinition}{\plaIteration=\plaGenericVector{i}\mtrans \in \plaIterationSpace}

\newcommand{\plaEquation}[1][i]{S_{#1}}
\newcommand{\plaEquationSpace}{\mset{S}}
\newcommand{\plaEquationDomain}[1][i]{\mset{I}_{#1}}
\newcommand{\plaEquationSpaceDefinition}{\plaEquationSpace = \{\plaEquation[0], \plaEquation[1], \dots\}}
\newcommand{\plaOperation}[1][i]{F_{#1}}
\newcommand{\plaOperationArity}[1][i]{1 \le j \le \text{arity}(\plaOperation[#1])}

\newcommand{\plaWriteIndexingFunctionMatrix}[1][i]{\mmat{P}_{#1}}
\newcommand{\plaWriteIndexingFunctionOffset}[1][i]{\mvec{f}_{#1}}
\newcommand{\plaWriteIndexingFunction}[1][i]{\plaWriteIndexingFunctionMatrix[#1]\plaIteration + \plaWriteIndexingFunctionOffset[#1]}

\newcommand{\plaReadIndexingFunctionMatrix}[1][{i, j}]{\mmat{Q}_{#1}}
\newcommand{\plaReadIndexingFunctionOffset}[1][{i, j}]{\mvec{d}_{#1}}
\newcommand{\plaReadIndexingFunction}[1][{i, j}]{\plaReadIndexingFunctionMatrix[#1]\plaIteration - \plaReadIndexingFunctionOffset[#1]}

\newcommand{\plaVariable}[1]{#1}
\newcommand{\plaVariableSpace}{\mset{X}}

\newcommand{\plaInputVariableSpace}{\plaVariableSpace_{\text{in}}}

\newcommand{\plaOutputVariableSpace}{\plaVariableSpace_{\text{out}}}

\newcommand{\plaInternalVariableSpace}{\plaVariableSpace_{\text{var}}}

\newcommand{\plaVariableAt}[2]{\plaVariable{#1}[#2]}

\newcommand{\plaEquationDefinition}{\plaEquation[i]: \plaVariableAt{x_i}{\plaWriteIndexingFunction} = \plaOperation[i](\dots, \plaVariableAt{y_{i, j}}{\plaReadIndexingFunction}, \dots) \quad \text{if } \plaIteration \in \plaEquationDomain[i]}

\newcommand{\plaTileSize}{p}

\newcommand{\plaTileSizesDimension}{\plaDimensionVector{\plaTileSize}{i}}

\newcommand{\plaIntra}{^j}
\newcommand{\plaInter}{^k}

\newcommand{\plaTileCount}{t}

\newcommand{\plaTileCountsDimension}{\plaDimensionVector{\plaTileCount}{i}}

\newcommand{\plaIntraIterationSpace}{\mset{J}}

\newcommand{\plaInterIterationSpace}{\mset{K}}

\newcommand{\plaTiledIterationSpace}{\mset{I}^*}

\newcommand{\plaTiledIntraScheduleVector}{\mvec{\lambda}\plaIntra}
\newcommand{\plaTiledInterScheduleVector}{\mvec{\lambda}\plaInter}
\newcommand{\plaTiledScheduleVector}{\mvec{\lambda^*}}
\newcommand{\plaTiledScheduleVectorDefinition}{\mvec{\lambda^*} = (\plaTiledIntraScheduleVector, \plaTiledInterScheduleVector)}
\newcommand{\plaScheduleOperationStartTime}[1][i]{\tau_{#1}}
\newcommand{\plaScheduleOperationExecutionTime}[1][i]{\delta_{#1}}

\newcommand{\plaScheduleInitiationInterval}{II}

\newcommand{\plaConditionSpace}{\plaEquationDomain}
\newcommand{\plaConditionSpaceMatrix}{\mmat{A}_i}
\newcommand{\plaConditionSpaceMatrixDefinition}{\plaConditionSpaceMatrix \in \mbaseSet{Z}^{m \times \plaDimension}}
\newcommand{\plaConditionSpaceOffset}{\mvec{b}_i}
\newcommand{\plaConditionSpaceOffsetDefinition}{\plaConditionSpaceOffset \in \mbaseSet{Z}^\plaDimension}

\newcommand{\plaConditionSpaceLongDefinition}{\plaConditionSpace = \msetBuilder{\plaIteration \in \plaIterationSpace}{\plaConditionSpaceMatrix\plaIteration \ge \plaConditionSpaceOffset}}
\newcommand{\plaConditionSpaceDefinition}{\plaConditionSpace \subseteq \plaIterationSpace}

\newcommand{\plaStorageLayout}{\mvec{s}}
\newcommand{\plaStorageLayoutOffset}{\alpha}

\newcommand{\plaAddressTranslation}{\mvec{m}}
\newcommand{\plaAddressTranslationOffset}{\mu}

\newcommand{\plaAddressTranslationDefinition}{\plaAddressTranslation_x\plaIteration+\plaAddressTranslationOffset_x = \begin{cases}%
    \plaStorageLayout_x(\plaReadIndexingFunction[{i_x, j_x}]) + \plaStorageLayoutOffset_x & x \in \plaInputVariableSpace\\
    \plaStorageLayout_x(\plaWriteIndexingFunction[i_x]) + \plaStorageLayoutOffset_x & x \in \plaOutputVariableSpace
\end{cases}}

\DeclareSIUnit\flops{FLOPS}
\DeclareSIUnit\flop{FLOP}
\DeclareSIUnit\luts{LUTs}
\DeclareSIUnit\ffs{FFs}

\DeclareSIUnit\ops{OPS}
\DeclareSIUnit\op{OP}

\begin{document}

\newcommand{\tikzpic}[3]{
    \tikzset{pics/#1/.style n args={#2}{code={#3}}}
}

\newcommand{\eg}{e.\,g.}
\newcommand{\ie}{i.\,e.}
\newcommand*\circled[1]{\tikz[baseline=(char.base)]{
            \node[shape=circle,draw,inner sep=1pt] (char) {#1};}}
\definecolor{cmark_green}{HTML}{00A64F}
\newcommand{\cmark}{\textcolor{cmark_green}{\ding{51}}}
\definecolor{xmark_red}{HTML}{ED1B23}
\newcommand{\xmark}{\textcolor{xmark_red}{\ding{55}}}
\definecolor{mmark_yellow}{HTML}{ED7A1B}
\newcommand{\mmark}{\textcolor{mmark_yellow}{\ding{51}}}

\newcolumntype{C}[1]{>{\centering\let\newline\\\arraybackslash\hspace{0pt}}m{#1}}

\newcommand{\secRef}[1]{Section~\ref{#1}}
\newcommand{\codeRef}[1]{Listing~\ref{#1}}
\newcommand{\figRef}[1]{Figure~\ref{#1}}
\newcommand{\tableRef}[1]{Table~\ref{#1}}
\setlength\arraycolsep{1pt}
\newcommand{\praEqRef}[1]{(Eq.~$S_{#1}$)}

\newcommand{\factor}[1]{\qty{#1}{\times}}
\newcommand{\factorRange}[2]{\qtyrange{#1}{#2}{\times}}

\newcommand{\memSize}[2]{#1$\times$\qty{#2}{\bit}}

\title{Mapping and Execution of Nested Loops on \\Processor Arrays: CGRAs vs. TCPAs}

\author{
	\IEEEauthorblockN{Dominik Walter, Marita Halm, Daniel Seidel, Indrayudh Ghosh,\\ Christian Heidorn, Frank Hannig, J\"urgen Teich\\}
	\IEEEauthorblockA{cs12-alpaca@fau.de\\
	Hardware/Software Co-Design, Department of Computer Science\\
	Friedrich-Alexander-Universit\"at Erlangen-N\"urnberg (FAU), Germany
			}
}

\maketitle

\begin{abstract}
Increasing demands for computing power also propel the need for energy-efficient SoC accelerator architectures.
One class of such accelerators are so-called processor arrays, which typically integrate a two-dimensional mesh of interconnected processing elements~(PEs).
Such arrays are specifically designed to accelerate the execution of multidimensional nested loops by exploiting the intrinsic parallelism of loops.
Moreover, for mapping a given loop nest application, two opposed mapping methods have emerged: Operation-centric and iteration-centric.
Both differ in the granularity of the mapping.
The operation-centric approach maps individual operations to the PEs of the array, while the iteration-centric approach maps entire tiles of iterations to each PE.
The operation-centric approach is applied predominantly for processor arrays often referred to as Coarse-Grained Reconfigurable Arrays~(CGRAs), while processor arrays supporting an iteration-centric approach are referred to as Tightly-Coupled Processor Arrays~(TCPAs) in the following.
This work provides a comprehensive comparison of both approaches and related architectures by evaluating their respective benefits and trade-offs.
We analyzed five toolchains for loop mapping on CGRAs and TCPAs, evaluating as well qualitative factors (\eg, intuitiveness, flexibility) as quantitative metrics (power, performance, area).
As a result, it is shown that for an equal number of PEs, both architectures offer distinct advantages:
The simpler structured CGRAs offer a better area cost~(\factor{6.26}), and a lower power consumption~(\factor{1.69}) when implemented on an FPGA, while TCPAs dominate in terms of achievable performance and efficiency, outperforming CGRAs in all tested benchmarks by up to \factor{19}.
This suggests that CGRAs are particularly suitable for area-constrained applications, while TCPAs excel in performance-critical scenarios by exploiting multicycle operations as well as multi-level parallelism in higher-dimensional loop nests.
\end{abstract}


\section{Introduction}
    The escalating demand for computing power has driven the evolution of diverse compute architectures, each tailored to maximize performance, efficiency, or scalability.
    A prominent approach is organizing many processing elements (PEs) into a two-dimensional grid, known as a processor array.
    This class can be further distinguished between arrays of complex, full-fledged processor cores and arrays in which the PEs act more as compute units with minimal programmability.
    Processors of the first class are commonly known as manycores~\cite{TeichHerkersdorf2022Book, SCC} or Massively Parallel Processor Arrays~(MPPAs)~\cite{kalray} and target general-purpose applications written in typical parallel programming models such as, for example, OpenCL.
    In this work, we focus on the alternative: Arrays of tiny PEs that utilize a more fine-grained mapping approach.
    Unlike GPUs that exploit vectorized data processing over multiple programs in SPMD mode, or application-specific processors such as tensor cores and AI accelerators, these arrays offer individually programmable and locally register-to-register communicating PEs to efficiently support the execution of loop nets with loop-carried dependencies.
    Therefore, processor arrays are particularly well-suited for executing multidimensional nested loops in parallel such as matrix computations and linear algebra problems in general.
    Such applications contain a substantial amount of inherent parallelism that can be exploited across multiple PEs.
    However, the process of automatically mapping such loops onto an array is complex and requires sophisticated mapping strategies.
    Recent years have seen a tremendous boost in the development of processor arrays specifically designed to target loop nests, leading to a diversity of architectures, mapping approaches, and toolchains.
    Out of this research, two principally different mapping philosophies have emerged:
    \begin{enumerate}
        \item \textbf{Operation-centric mapping:}
            A multidimensional loop nest is sequentially executed except of the innermost loop.
            The operations and data dependencies of the loop body are captured in a data flow graph (DFG).
            Each PE is then assigned a set of operations (nodes in the DFG), while data dependencies (edges) are mapped onto the interconnections between PEs.
            A resulting sequence of operations per PE is then generated, loaded, and repeatedly executed in such a way that the execution of iterations of the innermost loop can overlap in time.
        \item \textbf{Iteration-centric mapping:}
            The mapping granularity extends beyond individual operations to entire tiles of iterations.
            Here, the multidimensional iteration space is divided into congruent tiles of iterations with all iterations within a tile being assigned for execution by a single PE.
            Hence, parallelism can be exploited potentially in multiple loop dimensions while preserving the inherit data locality of loops.
    \end{enumerate}
    Architectures supporting operation-centric mappings are commonly referred to as Coarse-Grained Reconfigurable Arrays (CGRAs)~\cite{26_CGRA_Overview,24_CGRA_Taxonomy}, while those following an iteration-centric mapping are called Tightly-Coupled Processor Arrays (TCPAs)~\cite{KisslerTeich2006IEEEFPT, HannigReiche2014ACMTECS, TeichWitterauf2022Book} in the following.
    Both approaches are targeting the acceleration of multidimensional loop programs, with authors claiming high speedups, low hardware costs, and low power consumption.
    This paper aims to evaluate the pros and cons of both mapping strategies and related architectures by providing a comparative analysis of CGRAs and TCPAs.
    Although both classes of architectures have been extensively studied, this paper is, to the best of our knowledge, the first to provide an in-depth comparison of both approaches by:
    \begin{itemize}
        \item Describing the essentials of the architecture classes, mapping approaches, and available toolchains for CGRAs and TCPAs in detail in \secRef{sec:cgra} and \secRef{sec:tcpa}.
        \item Examining qualitative factors, \ie, intuitiveness, robustness, correctness, scalability, flexibility, and limitations, of available CGRA and TCPA toolchains in \secRef{sec:quali}.
        \item Conducting a quantitative comparison of power, performance, and area (PPA) for both architecture classes in \secRef{sec:quant} for a set of loop nest benchmarks.
        \item Discussing resulting trade-offs and limitations of both classes of architectures and mapping approaches in \secRef{sec:discussion} and \secRef{sec:conclusion}.
    \end{itemize}
\section{Coarse-Grained Reconfigurable Arrays}
\label{sec:cgra}
    Coarse-grained reconfigurable arrays~(CGRAs) were first presented in the 1990s, and since then, many CGRA architectures have been developed by both industry and academia, see, \eg,~\cite{26_CGRA_Overview,24_CGRA_Taxonomy}.
    \Citeauthor{26_CGRA_Overview}~\cite{26_CGRA_Overview} state that most architectures are academic projects, but they have also found commercial usage~\cite{25_SamsungTV}.
    In the following, we will first introduce the architecture of CGRAs in \secRef{sec:cgra:arch} and discuss the mapping approach in \secRef{sec:cgra:mapping}.

        \begin{figure*}
            \centering
            \begin{minipage}[c]{0.5\textwidth}%
            \resizebox{1.0\textwidth}{!}{
                \includegraphics[width=0.5\textwidth]{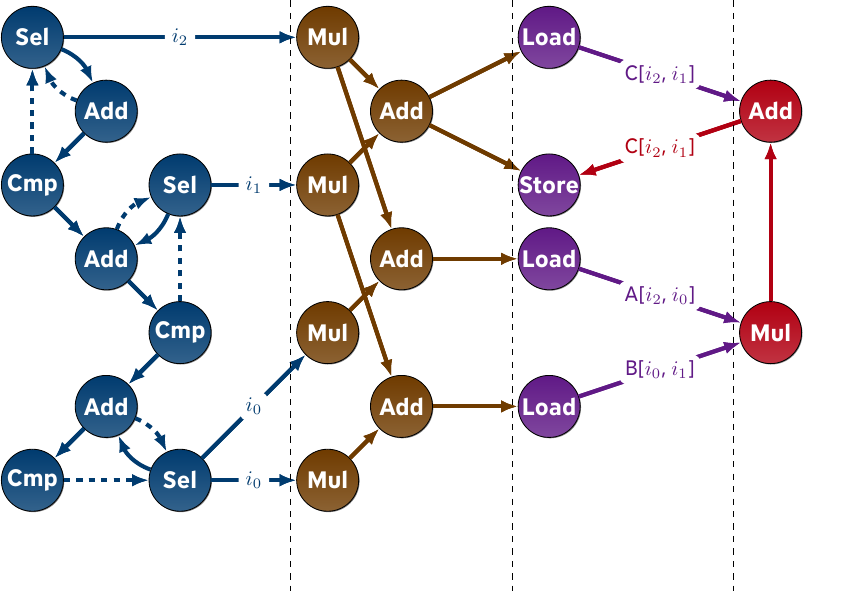}
            }
            \end{minipage}%
            \begin{minipage}[c]{0.5\textwidth}%
            \resizebox{1.0\textwidth}{!}{
                \includegraphics[width=0.5\textwidth]{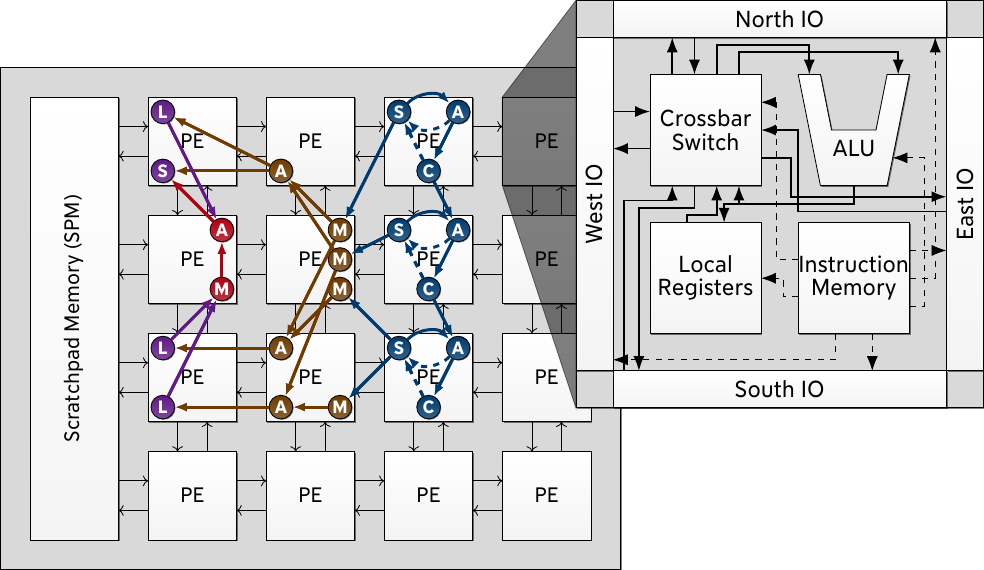}
            }
            \end{minipage}
            \caption{%
                Operation-centric mapping approach of Coarse Grained Reconfigurable Arrays.
                On the left, a simplified data flow graph~(DFG) of a matrix multiplication is shown.
                The nodes representing operations are grouped into indices computation~(blue), address computation~(brown), memory access~(purple) and finally multiply and accumulate operations~(red).
                Edges denote the data dependencies.
                This DFG is mapped onto the 4$\times 4$ CGRA architecture shown on the right.
                Each PE contains a functional unit (FU), \eg, an ALU, a local register file, switches, and a configuration memory, adapted from~\cite{2_HyCUBE}.
            }
            \label{fig:mainFigCgra}
        \end{figure*}

    \subsection{Architecture}
    \label{sec:cgra:arch}
        According to~\cite{19_CGRA}, a typical CGRA architecture consists of a network of interconnected PEs arranged in a two-dimensional grid, as shown in Figure \ref{fig:mainFigCgra}~(right).
        To keep the hardware simple and modular, each PE contains one functional unit~(FU), a set of local registers potentially arranged in a register file, a crossbar switch, and an instruction memory.
        The FU usually supports different arithmetic, logic, and memory operations at the word level.
        The local registers are used as temporary data storage for intermediate results.
        The crossbar connects the PE with its adjacent neighbors, enabling data transfer between neighboring PEs in a single cycle.
        The operation performed by the FU and the routing for the crossbar can be configured at the granularity of clock cycles.
        The instruction memory can store a sequence of predetermined per-cycle configurations to execute one loop iteration.
        For example, a configuration can specify that a FU of a PE performs an addition in one cycle, and that the crossbar forwards the result to a neighboring PE in north, east, south, or west direction.
        In the next cycle, the PE can perform a multiplication, storing the result in a local register.
        Such a predetermined sequence of configurations is repeated for each iteration of a given loop nest.
        Most CGRAs also support conditional execution by predication, \ie, the execution of some instructions is masked by a predication bit that was set by a conditional instruction.
        According to \figRef{fig:mainFigCgra}, typically only a subset of PEs has a direct access to an attached on-chip scratchpad memory (SPM) that can buffer input and output locally.
        Moreover, because only neighboring PEs can read or write data within one clock cycle, transferring data to a PE further away usually requires multiple cycles, while the intermediate PEs are then occupied for communication.
        HyCUBE~\cite{2_HyCUBE,WangKMMP19} alleviates these issues by a reconfigurable interconnect with single-cycle multi-hop connections.
    \subsection{Mapping}
    \label{sec:cgra:mapping}
        CGRAs are designed to accelerate nested loops by utilizing an operation-centric mapping approach.
        The operations and data dependencies of the loop body given from a multidimensional nested loop nest, specified, \eg, by a C/C++ program, are captured in a DFG $(V, E)$ in which a node $v_i \in V$ represents an operation, and an edge $(v_i, v_j) \in E$ represents a dependency between nodes $v_i$ and $v_j$.
        The mapping process of such a DFG onto a CGRA can be summarized as
        a) binding, \ie, assigning each node $v_i$ a PE $\beta(v_i)$,
        b) scheduling, \ie, assigning each node $v_i$ a start time $\tau(v_i)$, and
        c) routing, \ie, assigning each edge $(v_i,  v_j)$ a route connecting the PEs $\beta(v_i)$ and $\beta(v_j)$ such that the data arrives exactly at the right cycle at the FU of the target PE.
        This is achieved by an allocation of $r_{i, j}$ register slots, \ie, a register at a certain time, that must suffice $\tau(v_i) + d_i + r_{i, j} = \tau(v_j)$, where $d_i$ denotes the latency of the node (operation) $v_i$.
        The DFG describes the data dependencies of just one single loop iteration that is mapped and scheduled on the CGRA, in which node $v_i$ of loop iteration $i_2$ is to be computed at time $\tau(v_i) + \II \cdot i_2$, where $\II$ denotes the initiation interval.
        Since it is possible that there is a node $v_i$ that is planned after the next iteration has already been started, \ie, $\tau(v_i) \ge \II$, the execution of multiple iterations may overlap in time.
        Overall, the mapping process aims to minimize the $\II$ as it directly reflects the overall latency of the resulting loop nest execution.
        While there exists a wide variety of approaches and toolchains in the literature, we give in the following a general overview on the DFGs of multidimensional loop programs.
        Specific toolchains will be introduced later in \secRef{sec:cgra:tools}.

        Example.
        consider \figRef{fig:mainFigCgra}.
        Shown is a simplified but representative DFG of a typical 3-dimensional loop nest for computing a matrix-matrix multiplication.
        Each node denotes one operation, and the edges show the dependencies between the operations.
        For the execution of each iteration of the 3-dimensional loop nest with index vector $(i_0, i_1, i_2)$, four types of computations are involved:
        {(a)} Determination of the current loop indices---the corresponding operations are shown in \figRef{fig:mainFigCgra} on the left.
        Each loop index computation requires three operations, \eg, to compute the innermost loop (index $i_2$), a \texttt{Sel}, \texttt{Add}, and \texttt{Cmp} operation is needed.
        \texttt{Sel} is a multiplex operation that uses the result of the \texttt{Cmp} instruction to either forward the output of the \texttt{Add} operation or zero.
        The \texttt{Cmp} compares the result of the \texttt{Add} operation, which increments the current loop index, against a predefined constant, here, the loop bound.
        Note that the data dependencies towards the \texttt{Sel} operations are inter-iteration dependencies.
        This effectively implements a cyclic accumulator.
        Furthermore, the result of the \texttt{Cmp} operation can also be used as an addend for the \texttt{Add} operation of the next loop index.
        Therefore, the second level (index $i_1$) is only incremented once the first one reaches its loop bound, implementing a two-dimensional loop counter.
        This can be repeated for a third outer dimension (index $i_0$) as shown in \figRef{fig:mainFigCgra}.
        {(b)} Then, once the loop indices have been properly determined for the current iteration $(i_0, i_1, i_2)$, the addresses of the matrix elements that are to be accessed in this iteration must be computed.
        This is done by multiplying the loop indices with fixed strides and adding the results together.
        {(c)} Afterward, the computed addresses are used to load the inputs and store the output~(see the memory access section in \figRef{fig:mainFigCgra}).
        A restriction here is that in contrast to the other operations, the corresponding \texttt{Load} and \texttt{Store} operations cannot be executed on all PEs, but only on those PEs that have access to the SPM, which are, as shown in \figRef{fig:mainFigCgra}, only the border PEs.
        {(d)} Only then can the \texttt{Mul} and \texttt{Add} operations, forming the only computational part of the loop nest, \ie, one partial product, be computed before the result is written back by a \texttt{Store} operation.
        By studying the above DFG, we can already observe some interesting properties of CGRAs and related operation-centric mappings.
        First, note that the DFG contains multiple performance-constraining cycles, \eg, \texttt{Sel}$\,\to\,$\texttt{Add}$\,\to\,$\texttt{Cmp}$\,\to\,$\texttt{Sel}, inside the indices computation.
        As a consequence, the \texttt{Sel} operation of the next iteration cannot be started before the \texttt{Cmp} and \texttt{Add} operations of the previous iteration are completed.
        Thus, the cycle length determines a minimal possible \II, called the \emph{recurrence minimum initiation interval}~(RecMII).
        Also, the minimal possible \II may be further limited by a \emph{resource minimum initiation interval}~(ResMII).
        For example, given a CGRA with 9~PEs, the actual minimal possible \II is 3, because with $\II = 2$, each iteration would only allow for $9 \cdot 2 = 18$ nodes to be scheduled.
        This is due to the massive overhead for computing the indices and addresses.
        For the considered loop nest example, \ie, a loop body consisting of only a single MAC computation, the resulting DFG consists of a total of 22 nodes to be executed on the PEs per iteration.
        Note that this example is still highly simplified and real-world DFGs are much more complex, but similar in structure.
        The merits and drawbacks of such operation-centric mapping approaches are analyzed in \secRef{sec:quali} and \secRef{sec:quant}.
    \subsection{Tools}
    \label{sec:cgra:tools}
        Many different CGRA architectures have been proposed over the years, and some related toolchains are also publicly available.
        This paper selects four representative toolchains for analysis and comparison with TCPA approaches, including CGRA-Flow~\cite{4_OpenCGRA}, Morpher~\cite{6_MorpherWOSET}, Pillars~\cite{guo-pillars-woset2020}, and CGRA-ME~\cite{RaghebWWBRYA24}, which are briefly introduced in the following.
        \subsubsection{CGRA-Flow} \label{sec:cgra:tools:cgraflow}
            CGRA-Flow~\cite{4_OpenCGRA}, also known as Open\-CGRA, is a toolchain for the compilation, exploration, synthesis, and development of CGRA architectures~\cite{4_OpenCGRA}.
            It is open-source and available on GitHub\footnote{\url{https://github.com/tancheng/CGRA-Flow}}.
            CGRA-Flow has a GUI for visualizing input, output, and intermediate results.
            As input, users describe or select a loop program written in C/C++.
            CGRA-Flow supports an operation-centric mapping of up to two innermost loop nests with control flow in the loop body or up to three innermost loop nests without any control flow in the loop body onto the user-specified CGRA.
            Within the GUI, users can configure a CGRA architecture instance by selecting the number of PEs, the number of operations mapped to one PE, the size of the memory buffer, the operation types that the PE can execute, the connections to neighboring PEs, and the disablement of entire PEs.
            Note that each PE can only perform single-cycle operations.
            Compiling the user-given loop and generating the corresponding DFG is performed using LLVM's~\cite{LLVM} intermediate representation 
            to extract the operations and the dependency between operations.
            The generated DFG can be visualized in the GUI.
            Before starting the mapping phase, the user can select between two mapping algorithms, called \emph{exhaustive} and \emph{heuristic}.
            While the exhaustive algorithm checks all possible mappings for one given initiation interval \II, the heuristic approach starts with a minimal initiation interval and iteratively increments it until a mapping with the lowest cost (based on a heuristic function) for the current initiation interval \II has been found.
            The resulting mapping is then visualized in the GUI.
            After mapping, the user can generate Verilog via PyMTL to run various tests on the architecture~\cite{4_OpenCGRA} and to estimate the area and power of PEs and on-chip memory.
        \subsubsection{Morpher}\label{sec:cgra:tools:morpher}
            Morpher is an integrated compilation and simulation toolchain~\cite{6_MorpherWOSET} available on GitHub\footnote{\url{https://github.com/ecolab-nus/morpher}}.
            As input, the user provides a description of the target CGRA architecture and a loop program written in C/C++ that should be mapped onto the target architecture.
            The DFG generator begins by extracting the innermost loop of the program and generating the corresponding DFG using LLVM~\cite{LLVM}.
            It offers three schemes for handling loop control flow in a CGRA, influencing DFG generation: \emph{partial predication}, \emph{full predication}, and \emph{dual-issue}~\cite{6_MorpherWOSET}, described in detail in~\cite{8_Branch_aware_loop_mapping}.
            Partial predication maps the if-part and else-part operations to different PEs, adding a \emph{select} node if both parts update the same variable.
            Full predication schedules both parts using the same variable to the same PE, with one operation executed per cycle, avoiding the need for a select node.
            Dual-issue merges both operations into one DFG node, scheduling them simultaneously but only executing one at run-time.
            In this work, we only consider partial predication because it was the most reliably supported mapping.
            The resulting DFG and data layout for input/output variables on the SPM are then used by the CGRA mapper to find a valid mapping using three algorithms: PathFinder~\cite{McMurchieE95}, Simulated Annealing~\cite{32_SA}, and LISA~\cite{LiMitra2022HPCA}.
            The mapping can then be verified against automatically created test data~\cite{3_MorpherCODAI} by simulating the execution using a simulator that models a CGRA with FUs, registers, multiplexers, and memory banks~\cite{6_MorpherWOSET}, supporting variations of HyCUBE.
        \subsubsection{Pillars}\label{sec:cgra:tools:pillars}
            Pillars is an open-source CGRA design toolchain based on Scala and Chisel~\cite{guo-pillars-woset2020}.
            The toolchain is publicly available on GitHub\footnote{\url{https://github.com/pku-dasys/pillars}}.
            It has been designed as a tool for conducting design space explorations and further hardware optimizations of CGRAs.
            The user must provide a Scala-based architecture description of the CGRA, which is then systematically converted by Chisel into a synthesizable Verilog description of the CGRA.
            The design can be synthesized for FPGAs to determine the performance, area and power consumption of the CGRA.
            In contrast to the other toolchains, the Pillars toolchain does not support any automated DFG generation from source code.
            Thus, the user must already provide the application as a DFG.
            Pillars offers two mapping algorithms, an ILP mapper and a Heuristic Search mapper.
            The ILP mapper is slow but succeeds more frequently than the Heuristic Search mapper.
            Therefore, the ILP mapper is used for all loop kernels in this work.
            The resulting mapping can then be either simulated on the RTL by Verilator, or used to determine the performance, area and power estimate for the execution on the actual hardware.
        \subsubsection{CGRA-ME}\label{sec:cgra:tools:cgrame}
            CGRA-ME is an open-source toolchain for modeling and exploration of CGRAs~\cite{RaghebWWBRYA24}.
            It is currently available in its second version~(first release~\cite{17_CGRA-ME}) and supports end-to-end CGRA compilation and simulation with RTL code generation.
            The toolchain is open-source and can be downloaded from the project's website\footnote{\url{https://cgra-me.ece.utoronto.ca/download/}}.
            It uses LLVM to extract a DFG from a given C/C++ source code.
            However, it does not support any nested loops as inputs, but although it supports partial predication, no support for conditional code was available.
            CGRA-ME maps the extracted DFG onto a target CGRA, which is specified by a provided architecture description.
            It offers to choose between three different operation-centric mapping approaches.
            First, CGRA-ME supports an ILP-based mapping that finds an optimal mapping but is very slow.
            Additionally, a heuristic approach reduces the search space of the ILP.
            Finally, CGRA-ME also includes a so-called clustered mapper that incorporates a simulated-annealing approach~\cite{32_SA} utilizing both QuickRoute~\cite{LiE04} and PathFinder~\cite{McMurchieE95}.
            After the mapping, CGRA-ME produces a Verilog description of the given architecture and a bitstream containing the configuration of the CGRA.
            However, simulation without additional external toolchains is not available.
\section{Tightly-Coupled Processor Arrays}
\label{sec:tcpa}
    In contrast to CGRAs, so-called Tightly-Coupled Processor Arrays~(TCPAs)~\cite{KisslerTeich2006IEEEFPT, HannigReiche2014ACMTECS, TeichWitterauf2022Book} are designed such to support an iteration-centric mapping approach.
    TCPAs are constructed to enable the parallel execution of multidimensional loop nests that are expressed using a strict mathematical model related to polyhedral recurrence equations called \textit{Piecewise Regular Algorithms}~(PRAs)~\cite{Tei93, TeichT93}, see \secRef{sec:tcpa:pra}.
    This specification allows for the most natural description and mapping of loops~\cite{WitteraufWHT21, WalterWT20}.
    For example, contrary to a loop nest in an iterative language such as C/C++, there is no implied order of iteration or operation execution existing at all.
    In the following, we give first in \secRef{sec:tcpa:architecture} a short overview of the hardware architecture followed by further details on the iteration-centric mapping approach.

\begin{figure*}
    \centering
    \resizebox{1.0\textwidth}{!}{
    \begin{tikzpicture}
        \node[inner sep = 0] (tcpa) at (0, 0) {\includegraphics[width=0.5\textwidth]{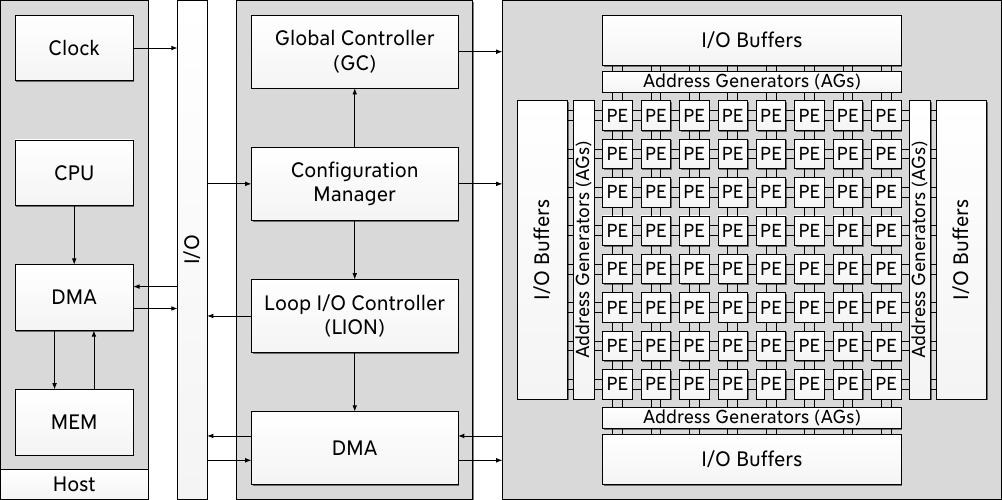}};
        \node[inner sep = 0] (pe) at (0.5\textwidth+1cm, 0) {\includegraphics[width=0.5\textwidth]{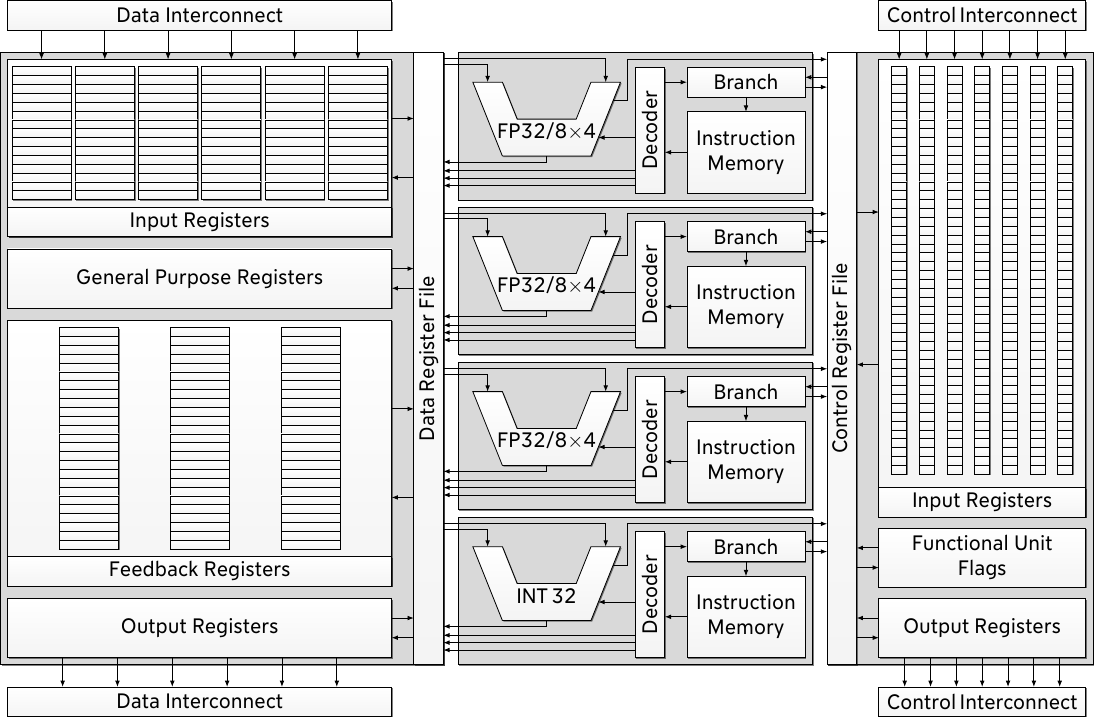}};

        \draw[fill=black, opacity=0.5] ([xshift=-1.19cm, yshift=-0.92cm]tcpa.north east) -- (pe.north west) -- (pe.north east) -- (pe.south east) -- (pe.south west) -- ([xshift=-1.19cm, yshift=-1.20cm]tcpa.north east) -- cycle;
        \node[inner sep = 0] (pe) at (0.5\textwidth+1cm, 0) {\includegraphics[width=0.5\textwidth]{alpaca_pe.pdf}};
    \end{tikzpicture}
    }
    \caption{
        Architecture of an $8\times 8$ TCPA (left) and the OIP-based~\cite{BrandTeich2017IEEEMCSOC} PE architecture (right) from \cite{alpaca}.
        The array is surrounded by 4 I/O buffers with address generators and has peripheral controllers shown left to the array.
        Each PE has a data and control register file and may have multiple functional units.
    }
    \label{fig:tcpaArch}
\end{figure*}

    \subsection{Hardware Architecture}
    \label{sec:tcpa:architecture}
        Both CGRA and TCPA architectures feature a two-dimensional array of small programmable PEs with a configurable circuit-switched interconnect, as shown in \figRef{fig:tcpaArch}.
        However, TCPAs differ significantly in their design and capabilities.
        Unlike CGRAs, each PE of a TCPA may be configured to include not only one but multiple parallel functional units (FUs).
        Different to a VLIW organization that can suffer from low code densities, each PE adheres to the principle of orthogonal instruction processing~(OIP), see~\cite{BrandTeich2017IEEEMCSOC}.
        In principle, each FU has its own instruction memory, branch unit, and program counter, allowing it to run independent micro-programs but sharing flags and registers with all other FUs.
        Such micro-programs consist of FU-specific instructions and branch instructions.
        While each FU instruction refers to a certain operation $\plaOperation$, branch instructions handle conditional jumps within the program.
        Operand dependencies are managed through a local data register file, with two source operands and one destination operand specified in each instruction, while the control signals used for the branch instructions originate from a control register file.
        The latter is connected to a Global Controller (GC) that coordinates and synchronizes all PEs, generating the compiler-generated schedule once for the entire array of PEs.
        The LION I/O controller~\cite{WalterT21} manages DMA transfers, fetching input and pushing output data between external memory and the four I/O buffers surrounding the array.
        Address calculations are all performed by programmable address generators within each memory bank.
        This approach fully relieves PEs from handling loop indices computations and address calculations.
    \subsection{Piecewise Regular Algorithms~(PRA)}
    \label{sec:tcpa:pra}
        In a PRA~\cite{Tei93, TeichT93}, a multidimensional for-loop with $\plaDimension$ nests is formed by an $\plaDimension$-dimensional polyhedral iteration space $\plaIterationSpaceDefinition$.
        Each iteration index $\plaIterationLongDefinition$ denotes a single iteration of the loop nest.
        Input data, output data, and computations are defined on a set of variables $\plaVariable{x} \in \plaVariableSpace$.
        Each instance of a variable is indexed by $\plaVariableAt{x_i}{\plaIteration}$, \ie, $\plaVariable{x_i}$ at iteration index $\plaIteration$.
        The loop nest itself is described by a set of quantized equations $\plaEquationSpaceDefinition$ that define data dependencies between variables within the iteration space $\plaIterationSpaceDefinition$.
        These equations are defined as follows, see, \eg,~\cite{Tei93, TeichT93}:
        $$\plaEquationDefinition$$
        For all $\plaIteration \in \plaConditionSpaceDefinition$, the variable $\plaVariable{x}_i$ is defined as the result of applying the function $\plaOperation$ on a list of variables $\plaVariable{y}_{i, j}$ with $\plaOperationArity$.
        Affine indexing functions $\plaWriteIndexingFunction$ and $\plaReadIndexingFunction$ provide the indices with which the variables are accessed.
        Furthermore, we name those variables that are only read but not defined as the set of \textit{input variables} $\plaInputVariableSpace$ and those variables that are only defined but not used as the set of \textit{output variables} $\plaOutputVariableSpace$ and all others as the set of \textit{internal variables} $\plaInternalVariableSpace$.
        In a PRA, the indexing functions of the internal variables are further restricted to simple translations, \ie, $\plaWriteIndexingFunctionMatrix$ and $\plaReadIndexingFunctionMatrix$ denote identity matrices.
        Each equation $\plaEquation$ is only defined within its domain, \ie, a subspace of the iteration space.
        The domain is given by a \textit{condition space} $\plaConditionSpace$ that can usually be described by a system of linear inequalities, thus, $\plaConditionSpaceLongDefinition$ with $\plaConditionSpaceMatrixDefinition, \plaConditionSpaceOffsetDefinition$.
        As an example, \figRef{fig:praCode} lists the equations of a PRA implementing a matrix multiplication computing $\mmat{C} = \mmat{A} \cdot \mmat{B}$ with $\mmat{A}, \mmat{B}, \mmat{C} \in \mbaseSet{Z}^{N \times N}$.
        $\mmat{A}$ is read-in at $i_1 = 0$~\praEqRef{1, a}, and propagated along $i_1 > 0$~\praEqRef{1, b}.
        Similarly, $\mmat{B}$ is read-in at $i_0 = 0$~\praEqRef{2, a}, and propagated along $i_0 > 0$~\praEqRef{2, b}.
        Thus, each iteration has access to a single element of $\mmat{A}$ and a single element of $\mmat{B}$ that are multiplied~\praEqRef{3}.
        The result is propagated along $i_2$~\praEqRef{4, a}, where it is accumulated in every step~\praEqRef{4, b}.
        The final matrix elements are computed at $i_2 = N - 1$ and written to the output matrix $\mmat{C}$~\praEqRef{5, \mmat{C}}.
        Now, instead of performing an operation-centric mapping, TCPAs support an iteration-centric mapping that starts with a partitioning~(or tiling) step and a subsequent scheduling step that will be explained in the following together with a number of additional steps like register binding, code generation, I/O buffer allocation and, finally, the generation of the final TCPA configuration.
\begin{figure}
    \definecolor{darkGray}{RGB}{71, 79, 82}
    \centering
    \begin{align*}
        \nonumber S_{1,\text{a}}\colon & a[\mvec{i}] = \mmat{A}[i_{0},i_{2}]  \text{ if }  i_1 = 0\\
        \nonumber S_{1,\text{b}}\colon & a[\mvec{i}] = a[i_{0},i_{1}-1,i_{2}]  \text{ if }  i_1 > 0\\
        \nonumber S_{2,\text{a}}\colon & b[\mvec{i}] = \mmat{B}[i_{2},i_{1}]  \text{ if }  i_0 = 0\\
        \nonumber S_{2,\text{b}}\colon & b[\mvec{i}] = b[i_{0}-1,i_{1},i_{2}]   \text{ if }  i_0 > 0\\
        \nonumber S_{3}\colon & p[\mvec{i}] = a[\mvec{i}] \cdot b[\mvec{i}]\\
        \nonumber S_{4,\text{a}}\colon & c[\mvec{i}] = p[\mvec{i}]  \text{ if }  i_2 = 0\\
        \nonumber S_{4,\text{b}}\colon & c[\mvec{i}] = c[i_{0},i_{1},i_{2}-1] + p[\mvec{i}]  \text{ if }  i_2 > 0\\
        \nonumber S_{5,\text{C}}\colon & \mmat{C}[i_{0},i_{1}] = c[\mvec{i}]  \text{ if }  i_2 = N - 1
    \end{align*}
    \caption{%
        A matrix multiplication $\mmat{C} = \mmat{A} \cdot \mmat{B}$ with $\mmat{A}, \mmat{B}, \mmat{C} \in \mbaseSet{Z}^{N \times N}$ expressed as PRAs with iteration space $\mset{I} = \{ (i_0, i_1, i_2)\mtrans \in \mbaseSet{Z}^3 \mid 0 \leq i_0, i_1, i_2 < N \}$.
        The notations $x[\mvec{i}]$ and $x[i_0,i_1,i_2]$ are equivalent.
    }
    \label{fig:praCode}
\end{figure}

        \begin{figure}
            \centering
            \resizebox{0.5\textwidth}{!}{
                \includegraphics[width=0.5\textwidth]{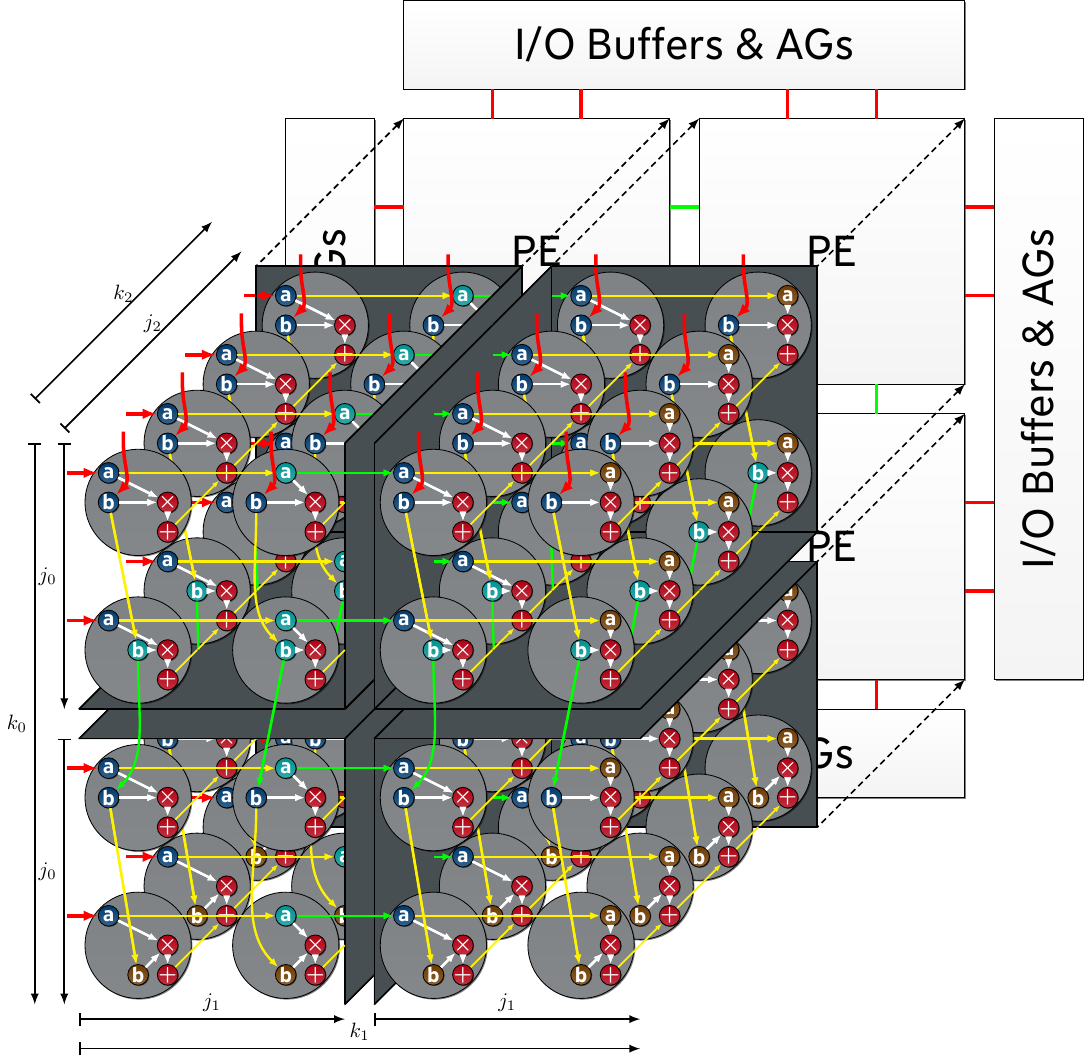}
            }
            \caption{%
                Simplified $4\times 4 \times 4$ iteration space of a matrix multiplication that is tiled into $2 \times 2 \times 1$ tiles and mapped onto a $2 \times 2$ PE array shown behind.
                Each gray circle denotes one iteration consisting of 4 operations as specified by the loop body.
                The edges denote data dependencies, whereas its color denote its type, \ie, input~(red), intra-iteration (white), inter-iteration intra-tile (yellow), or inter-iteration inter-tile (green).
                Although each iteration contains the same operations, the type of the data dependencies of the contained operations is different, which is reflected by both the position and color of the operation.
            }
            \label{fig:plaAlgo}
        \end{figure}

    \subsection{Partitioning}
    \label{sec:tcpa:partitioning}
        The iteration space $\plaIterationSpace$ is partitioned into $\plaTileCountsDimension$ rectangular tiles of size $\plaTileSizesDimension$.
        Each tile, a subset of the iteration space, is then mapped to a single PE.
        This approach follows a \textit{local sequential, global parallel}~(LSGP) strategy~\cite{Tei93, TeichT93, nelis1988}, where each PE implements in its instruction memory a sequential schedule of all iterations within its assigned tile, while all PEs run concurrently.
        Formally, the partitioned iteration space $\plaTiledIterationSpace$ is decomposed into an intra-tile space $\plaIntraIterationSpace$ and an inter-tile space $\plaInterIterationSpace$, where $\plaIntraIteration \in \plaIntraIterationSpace$ denotes the index vector of an iteration within a tile, and $\plaInterIteration \in \plaInterIterationSpace$ indicates the PE index responsible for executing that iteration.
        As an example, consider again the PRA of the matrix multiplication shown in \figRef{fig:praCode}.
        \figRef{fig:plaAlgo} illustrates the corresponding iteration space after partitioning in case of an input matrix size $N = 4$.
        Formally, the $4 \times 4 \times 4$ iteration space is divided into $t_0 \times t_1 \times t_2 = 2 \times 2 \times 1$ tiles, each of size $p_0 \times p_1 \times p_2 = 2 \times 2 \times 4$.
        As a result, each PE is assigned 16 iterations for execution.
        Each iteration (represented by gray circles) contains four operations: two \texttt{MOV}s for loading and propagating matrix elements $a \in \mmat{A} \in \mbaseSet{Z}^{4\times 4}$ and $b \in \mmat{A} \in \mbaseSet{Z}^{4\times 4}$, and two arithmetic operations to compute $ c = c + a \cdot b$.
        The arrows between the operations indicate data dependencies that can be further distinguished.
        Consider the definition of an equation $\plaEquation$.
        The result of operation $\plaOperation$ defines the variable $\plaVariable{x}_i$, while $\plaOperation$ itself uses a list of variables $\plaVariable{y}_{i, j}$ as inputs.
        This creates data dependencies among equations that can be classified as either intra-iteration, inter-iteration, or input/output if $\plaVariable{x}_i$ is in the output variable space ($\plaOutputVariableSpace$) or input variable space ($\plaInputVariableSpace$).
        Intra-iteration dependencies, shown by white arrows in \figRef{fig:plaAlgo}, occur within the same iteration when both $\plaWriteIndexingFunctionOffset$ and $\plaReadIndexingFunctionOffset$ are zero.
        Inter-iteration dependencies occur between neighboring iterations and are categorized as intra-tile (yellow) or inter-tile (green).
        Intra-tile dependencies happen within the same tile, while inter-tile dependencies occur across tiles, with each tile executed by a different PE.

    \subsection{Scheduling}
    \label{sec:tcpa:scheduling}
        The execution of a tile of iterations on a PE requires that each operation $\plaOperation$ as defined by an equation $\plaEquation$ is assigned a start time $\plaScheduleOperationStartTime$ that satisfies all intra-iteration data dependencies.
        In the example, each iteration includes four operations that must be mapped and scheduled on functional units within a PE, with each operation $\plaOperation$ characterized by an execution time $\plaScheduleOperationExecutionTime$ in clock cycles on the respective functional unit\footnote{Note that TCPAs naturally support multicycle FU operations.}.
        The execution of successive iterations within a tile then starts every initiation interval~$\plaScheduleInitiationInterval$ cycles.
        Finally, the scheduler must determine the order in which the iterations are started such that all inter-iteration data dependencies are satisfied.
        This so-called loop schedule is described by a linear schedule vector $\plaTiledScheduleVectorDefinition$, where the start time of each intra-tile iteration $\plaIntraIteration \in \plaIntraIterationSpace$ is given by $\plaTiledIntraScheduleVector\plaIntraIteration$, and the start time of each inter-tile iteration (and PE) $\plaInterIteration \in \plaInterIterationSpace$ is defined by $\plaTiledInterScheduleVector\plaInterIteration$.
        %
        Further details on the scheduling algorithms are available in~\cite{TeichTZ97, TeichT02}.
        Note that this approach even supports the scheduling of loop nests with at-compile time unknown loop bounds~\cite{WitteraufTeich2016ASAP, TanaseHannig2018ATECS}.

    \subsection{Register Binding}
    \label{sec:tcpa:registerBinding}
        Finally, with each iteration and operation scheduled, registers for data dependencies can be allocated and bound accordingly.
        To handle different data dependencies, each PE has a register file with specialized register types for each dependency type, detailed in the following.
        The mapping of the inputs and output dependencies is explained later in \secRef{sec:tcpa:ioAllocation}.
        \subsubsection{General Purpose Registers (RDs)}
            Single-word registers for intra-iteration or inter-iteration intra-tile dependencies.
            In both cases, the lifetime, \ie, the number of cycles between write and read, must be shorter than the initiation interval.
        \subsubsection{Feedback Registers (FDs)}
            Internal FIFOs for dependencies where the lifetime exceeds the initiation interval.
            These are typically inter-iteration intra-tile dependencies that are written multiple times before the first read, as visualized, for example, in \figRef{fig:plaAlgo}.
            FDs can also be used to map intra-iteration dependencies when needed.
        \subsubsection{Input/Output Registers (IDs/ODs)}
            For inter-iteration dependencies across tiles (inter-tile dependencies), communication between PEs is enabled via a configurable interconnect.
            Output registers act as sending ports, pushing data through the interconnect to a target PE, where it is received into an input FIFO.
            This FIFO can then be accessed by the receiving PE by reading from a corresponding input register.
            The interconnect is dynamically configured to create communication channels for managing inter-tile dependencies.
        \subsubsection{Virtual Registers (VDs)}
            As shown in \figRef{fig:plaAlgo}, an operation may generate intermediate results needed within the same iteration, the next iteration, and even in neighboring tiles, for which different register types must be used.
            Therefore, TCPAs contain virtual registers that allow a single instruction to broadcast a write to multiple target registers simultaneously.
    \subsection{Code Generation}
    \label{sec:tcpa:programGeneration}
        Due to the different condition spaces $\plaConditionSpace$, the operations that are executed in an iteration may not always be the same.
        Refer, again, to \figRef{fig:plaAlgo}.
        While each tile is similar, minor differences necessitate different programs for each PE, though in larger arrays, multiple PEs may share the same program.
        Thus, the compiler generates, for all possible combinations of operations within an iteration, a sequence of instructions that reflects the computed start times, $\plaScheduleOperationStartTime$.
        Afterward, it uses branch instructions to implement an iteration sequencing that matches the loop schedule $\plaTiledIntraScheduleVector$.
        It is even possible that the execution of iterations overlaps in time, for which the instruction sequences must also be folded.
        Importantly, PEs do not compute the control flow directly; instead, the branch instructions respond to control signals generated by the Global Controller~(GC) as shown in~\figRef{fig:tcpaArch}.
        These signals, calculated once and distributed via a control interconnect, are shared across all PEs.
    \subsection{I/O Buffer Allocation}
    \label{sec:tcpa:ioAllocation}
        While registers handle internal variables, input and output dependencies, represented by red arrows in \figRef{fig:plaAlgo}, need a different mapping.
        The address of each element of a multidimensional input/output variable $x$, stored as linearized data in external memory, is given by a storage layout $\plaStorageLayout_x$ and offset $\plaStorageLayoutOffset_x$.
        This can be combined with the indexing of the variable to create an address translation $\plaAddressTranslation_x$ with offset $\plaAddressTranslationOffset_x$: $$\plaAddressTranslationDefinition$$
        Note that on purpose, these computations are not handled by the PEs but by dedicated address generators~(AGs) configured to compute any given affine address pattern.
        Consider again \figRef{fig:tcpaArch}.
        Since the PEs lack direct memory access, only the border PEs have access to I/O buffers containing small memory banks holding input and output data.
        The AGs address these memory banks, after which data is forwarded to the PEs through input registers.
        LION~\cite{WalterT21} manages the data transfer between the TCPA and external memory, filling and clearing the I/O buffers as required by a given schedule vector $\plaTiledScheduleVector$ in time.
    \subsection{Configuration Generation}
        The mapping of the PRA onto a TCPA is, finally, represented by a configuration---a binary file that programs the TCPA by loading the micro-programs of each FU and setting various configuration registers for, \eg, the interconnect, the AGs, the GC, or the LION~\cite{WalterT21}.
        This configuration file, loaded at runtime, enables the TCPA to independently execute multidimensional loop nests without external control.
    \subsection{Tools}
        \begin{figure}
            \resizebox{0.5\textwidth}{!}{
                \includegraphics[width=0.5\textwidth]{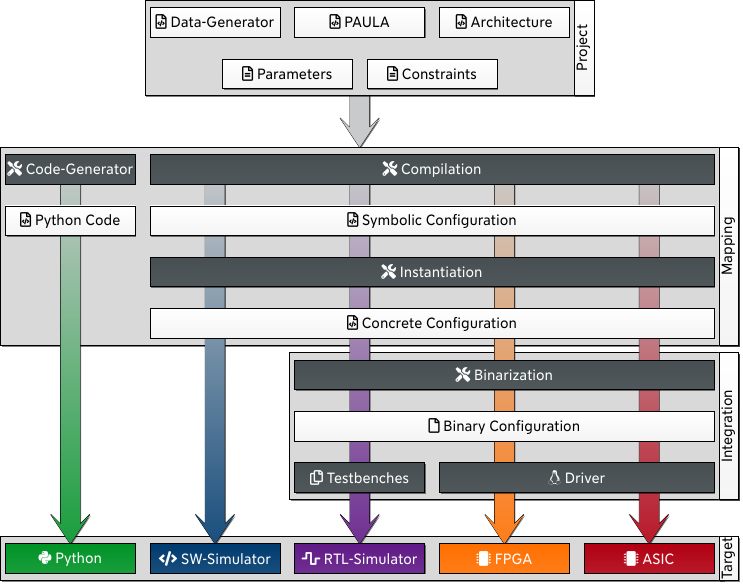}
            }
            \caption{Overview of the TURTLE toolchain for TCPAs.}
            \label{fig:turtle}
        \end{figure}
        \begin{lstlisting}[basicstyle=\tiny,style=my_PAULA_style, float, escapechar=|, label=listing:paula, captionpos=b,
          caption=PAULA code~\cite{PAULA} specifying computing a matrix multiplication.]
program matrix_multiplication {
  // In-/Output variables
  variable A 2 in float; variable B 2 in float; variable C 2 out float;
  // Internal variables
  variable a 2 float; variable b 2 float; variable p 2 float; variable c 2 float;
  // Parameters with values assigned at instantiation
  parameter N;
  // Loop with iteration space in brackets
  par (i_0 >= 0 and i_0 < N and i_1 >= 0 and i_1 < N and i_2 >= 0 and i_2 < N) {
    // Read matrix A and propagate along i_1
    a[i_0,i_1,i_2] = A[i_0,i_2] if (i_1 == 0);
    a[i_0,i_1,i_2] = a[i_0,i_1 - 1,i_2] if (i_1 > 0);
    // Read matrix B and propagate along i_0
    b[i_0,i_1,i_2] = B[i_2,i_1] if (i_0 == 0);
    b[i_0,i_1,i_2] = b[i_0 - 1,i_1,i_2] if (i_0 > 0);
    // Compute partial product
    p[i_0,i_1,i_2] = a[i_0,i_1,i_2] * b[i_0,i_1,i_2];
    // Accumulate along i_2
    c[i_0,i_1,i_2] = p[i_0,i_1,i_2] if (i_2 == 0);
    c[i_0,i_1,i_2] = p[i_0,i_1,i_2] + c[i_0,i_1,i_2 - 1] if (i_2 > 0);
    // Write matrix C
    C[i_0,i_1] = c[i_0,i_1,i_2] if (i_2 == N - 1);
  }
}
        \end{lstlisting}
        Unlike the range of toolchains available for CGRAs, we are only aware of one toolchain for TCPAs.
        \figRef{fig:turtle} illustrates the \textit{TCPA Utilities for Running and Testing Loop Executions}~(TURTLE) toolchain.
        First, the user provides a project.
        This contains the loop specified as a PAULA program~\cite{PAULA}, the target architecture, a data generator, parameters such as loop bounds and various mapping constraints.
        PAULA is a domain-specific programming language specifically designed to model PRAs (\secRef{sec:tcpa:pra}).
        For example, \codeRef{listing:paula} shows the PAULA code implementing the PRA of the matrix multiplication from \figRef{fig:plaAlgo}.
        The target architecture is specified in an XML file that details, \eg, the FUs within each PE, including FU instruction memory, PE registers, port connections, and I/O buffer sizes.
        Before compilation, the PAULA program can be transformed into equivalent Python code and, after execution, its results can be verified against reference values from the data generator.
        Next, the PAULA program is compiled into a so-called symbolic configuration~\cite{WitteraufWHT21, TanaseHannig2018ATECS}, primarily by generating a polyhedral syntax tree—a symbolic representation that specifies which operands are used in which operations within each FU across iterations.
        This requires computing a valid schedule and register binding, with the schedule achieved through symbolic modulo scheduling and regular RD registers bound using the left-edge algorithm.
        Compilation also involves partitioning the iteration space and identifying all data dependencies and required FD, ID, OD, and VD registers.
        This symbolic configuration is concretized during instantiation, where parameters such as problem size and PE count are set.
        Furthermore, configurations are generated for all AGs and the LION.
        Groups of PEs sharing the same FU programs are identified, and for each such so-called processor classes, the instantiator folds the polyhedral syntax tree, generating specific programs for all FUs in each class.
        Lastly, the control flow is extracted from the generated programs and converted into a GC configuration, completing the mapping phase.
        TURTLE supports various targets given a concrete configuration.
        For example, a cycle-accurate simulator can load the XML file, execute the loop, and verify results with the test data generator.
        For other targets, the XML file is converted to a binary file that can be loaded by either RTL testbenches or a TCPA driver for FPGA or ASIC targets.
        TURTLE also provides a highly generic VHDL codebase, from which concrete TCPA architectures can be generated and packaged into a dedicated IP core.

\section{Qualitative Evaluation}
\label{sec:quali}
    In this section, we conduct a qualitative evaluation of different features of the described compiler toolchains for CGRAs and TCPAs.
    This includes intuitiveness, robustness, correctness, scalability, flexibility, and limitations.
    An overview of the results is given in \tableRef{table:comp_feature}.

    \subsubsection{Intuitiveness}
        CGRA-Flow provides a GUI, visualizing input, output, and intermediate results, while the other toolchains only offer a rather simple commandline interface.
        Furthermore, most CGRA toolchains accept loops written in C/C++, which are commonly used programming languages.
        On the other hand, developers of applications for TCPAs need to specify their loop nests in a domain-specific polyhedral programming language, PAULA, which a user has to learn first.
        Thus, TCPAs are less intuitive to use compared to CGRAs.
    \subsubsection{Robustness}
        One crucial aspect of the mapping approach is its robustness, \ie, the ability to always find a valid mapping, if possible.
        Here, we observed that all toolchains require to perform manual optimizations on the code, otherwise the mapping often fails.
        Still, most toolchains were able to find valid mapping reliably.
        Only Pillars fails consistently.
    \subsubsection{Correctness}
        We found no obvious errors in the mapping process, however, the DFG generator of CGRA-ME tends to produce erroneous DFGs at times.
        Lastly, all toolchains offer a functional or cycle-accurate validation through simulation.
    \subsubsection{Scalability}
        The mapping complexity of loop nests to CGRAs scales with the number of PEs and the number of nodes in the DFG.
        This is due to the fact that each leads to an increase of the search space of possible mappings, which is explored by the CGRA toolchains.
        Only for CGRA-Flow, an increase in the number of PEs and DFG nodes does not noticeably affect the compilation time because the mapper only checks a single mapping per initiation interval $\II$.
        Thus, CGRA-Flow often finds a mapping within seconds if the heuristic algorithm is used.
        Nevertheless, CGRA-Flow still may take minutes to find a mapping if a generated DFG has many nodes (hundreds) or a CGRA has many PEs (64 or more), which is similar to the mapping times observed for Pillars and CGRA-ME.
        Only Morpher takes significantly longer---somewhere from minutes to hours, depending on the node count in a DFG and the PE count in a CGRA.
        For larger CGRAs, \ie, $8 \times 8$ PEs, and DFGs with more than 100 nodes, practically no CGRA toolchain provided a mapping in an affordable time of less than \qty{1}{\hour}.
        The obvious non-scalability of CGRA's mapping approaches is a known challenge, also mentioned in~\cite{36_survey_mapping_approach, 19_CGRA, 18_himap}.
        In contrast, the runtime of the mapping of a multidimensional loop nest onto TCPAs does neither increase with the number of PEs nor with the problem size, \ie, loop bounds, of the given loop nest, as the mapping and scheduling complex steps are performed symbolically, \ie, with parameterized loop bounds and array size, during compilation~\cite{WitteraufWHT21, TanaseHannig2018ATECS}.
        It only increases with the small number of equations (typically less than 10) within the PRA.
        In common is that for all toolchains, the problem size of the loop does not generally increase the mapping time.
    \subsubsection{Flexibility}
        Both CGRA and TCPA architectures are flexibly parameterized architectures.
        Hence, all compiler toolchains allow the user to configure the target architecture in terms of, \eg, the number of PEs or the supported operations, but there are subtle differences.
        In CGRA-Flow, the target architecture can be configured intuitively but is rather limited compared to the other toolchains, as neither multiple FUs per PE nor multicycle operations nor multi-hop connections are supported.
        From an application perspective, all toolchains but Pillars support the automated extraction of loops.
        However, only TURTLE and CGRA-Flow are able to map entire multidimensional loop nests.
        But the latter is restricted to only up to at most two nested loops, though.
        On the contrary, Morpher and CGRA-ME are limited to mapping only the innermost loop to the target CGRA.
        Note that although any multidimensional loop can be flattened to a single inner loop.
        This does, however, require predication, which was not yet available in CGRA-ME.
    \begin{table}
        \caption{Qualitative features of CGRA and TCPA toolchains.}
        \label{table:comp_feature}
        \centering
        \resizebox{0.5\textwidth}{!}{
            \definecolor{darkGray}{RGB}{71, 79, 82}
            \rowcolors{1}{darkGray!20}{darkGray!5}
            \setlength{\tabcolsep}{2pt}
            \begin{tabular}{|l|C{.15\linewidth}|C{.15\linewidth}|C{.15\linewidth}|C{.15\linewidth}|C{.15\linewidth}|}
                \rowcolor{darkGray}
                \color{white}{\textbf{Feature}} &
                \color{white}{\footnotesize {\textbf{CGRA-Flow\allowbreak\cite{4_OpenCGRA}}}} &
                \color{white}{\footnotesize {\textbf{Morpher \cite{6_MorpherWOSET}}}} &
                \color{white}{\footnotesize {\textbf{~~~Pillars~~~\allowbreak\cite{guo-pillars-woset2020}}}} &
                \color{white}{\footnotesize {\textbf{CGRA-ME \cite{RaghebWWBRYA24}}}} &
                \color{white}{\footnotesize {\textbf{TURTLE~\allowbreak\cite{WitteraufWHT21, TanaseHannig2018ATECS}}}} \\\hline

                \rowcolor{darkGray} \color{white}{\textbf{Intuitiveness}} &        &        &        &        &        \\\hline
                    $\bullet$ Graphical interface                          & \cmark & \xmark & \xmark & \xmark & \xmark \\
                    $\bullet$ Commandline interface                        & \cmark & \cmark & \cmark & \cmark & \cmark \\
                    $\bullet$ Commonly used language                       & \cmark & \cmark & \xmark & \cmark & \xmark \\
                \rowcolor{darkGray} \color{white}{\textbf{Robustness}}    &        &        &        &        &        \\\hline
                    $\bullet$ No manual optimization              & \xmark & \xmark & \xmark & \xmark & \xmark \\
                    $\bullet$ Reliable mapping success                     & \cmark & \cmark & \xmark & \cmark & \cmark \\
                \rowcolor{darkGray} \color{white}{\textbf{Correctness}}    &        &        &        &        &        \\\hline
                    $\bullet$ Simulation of mapping                        & \cmark & \cmark & \cmark & \xmark & \cmark \\
                    $\bullet$ Simulation statistics                        & \cmark & \xmark & \cmark & \xmark & \cmark \\
                    $\bullet$ Auto. test data generation                   & \xmark & \cmark & \xmark & \xmark & \xmark \\
                \rowcolor{darkGray} \color{white}{\textbf{Scalability}}    &        &        &        &        &        \\\hline
                    $\bullet$ Independent of \#Operations                 & \xmark & \xmark & \xmark & \xmark & \xmark \\
                    $\bullet$ Independent of \#Iterations                 & \cmark & \cmark & \cmark & \cmark & \cmark \\
                    $\bullet$ Independent of \#PEs                        & \mmark & \xmark & \xmark & \xmark & \cmark \\
                    $\bullet$ Independent of problem size                 & \cmark & \cmark & \cmark & \cmark & \cmark \\
                \rowcolor{darkGray} \color{white}{\textbf{Flexibility}}    &        &        &        &        &        \\\hline
                    $\bullet$ Generic \#PE                                 & \cmark & \cmark & \cmark & \cmark & \cmark \\
                    $\bullet$ Generic \#FU per PE                          & \xmark & \mmark & \cmark & \cmark & \cmark \\
                    $\bullet$ Generic interconnect                         & \cmark & \cmark & \cmark & \cmark & \cmark \\
                    $\bullet$ Generic operation latency                    & \xmark & \cmark & \cmark & \cmark & \cmark \\
                    $\bullet$ Generic hop length                           & \xmark & \cmark & \cmark & \cmark & \cmark \\
                    $\bullet$ Generic memory size                          & \cmark & \cmark & \cmark & \cmark & \cmark \\
                \rowcolor{darkGray} \color{white}{\textbf{Limitations}}    &        &        &        &        &        \\\hline
                    $\bullet$ Feature complete                             & \cmark & \cmark & \xmark & \mmark & \cmark \\
                    $\bullet$ Register-aware                               & \xmark & \cmark & \cmark & \cmark & \cmark \\\hline
            \end{tabular}
        }
    \end{table}
    \subsubsection{Limitations}
        With the exception of Pillars, which does not include a DFG generator, all tested toolchains are feature complete, \ie, they are able to generate a complete mapping to the target architecture.
        Note, however, that CGRA-Flow's mapping does not consider any PE register mapping.
        As a result, CGRA-Flow assumes an infinite number of registers within each PE.
        Additionally, CGRA-Flow does not consider the data layout within the larger memory buffer, or rather, it always assumes that the memory address starts at zero.
        A last limitation on mappability may be related to physical architecture limitations or constraints.
        Consider, \eg, the different register types of a TCPA that were introduced in \secRef{sec:tcpa}.
        Inter-iteration intra-tile dependencies within a loop are mapped to PE-local FIFO memories within the register file in each PE to be able to exploit data locality between multiple loop iterations which CGRAs cannot do.
        This reuse of data may, however, introduce architectural constraints due to the required length of these FIFOs, which correlates typically with the tile size after partitioning.
        As a result, the problem size of a loop can become limited by the available FIFO memory in a given TCPA architecture.
        Note that many real world problems can be decomposed into smaller subproblems that can be computed separately by individual calls to the accelerator~\cite{blu_paper}.
        In contrast, CGRAs cannot support any data locality at all, as they do not have local memory within the PEs.
        The only architectural constraint is the size of the peripheral memory around the array as it must be large enough to store the entire input and output data.
        However, note that TCPAs do not share this restriction, because they may refill the I/O buffers during runtime~\cite{WalterWT20, WalterT21}, which is not supported by any considered CGRA.

\section{Quantitative Evaluation}
\label{sec:quant}
    In this section, we evaluate different CGRA and TCPA processor array architectures quantitatively.
    Specifically, we compare the achievable Performance, Power, and Area~(PPA) of different architecture instances and mapping toolchains.

    \subsection{Performance}
    \label{sec:quant:performance}
        \begin{table}
    \caption{%
        Mapping results of benchmarks onto CGRAs and TCPAs.
    }
    \label{table:tcpa_cgra_mapping}
    \centering
    \resizebox{0.5\textwidth}{!}{
        \definecolor{darkGray}{RGB}{71, 79, 82}   
        \definecolor{darkRed}{RGB}{177, 0, 18}
        \definecolor{darkOrange}{RGB}{207, 112, 4}
        \rowcolors{1}{darkGray!20}{darkGray!5}
        \begin{tabular}{|l|c|c|c|c|c|c|c|}
          \hline 
 \rowcolor{darkGray} \color{white}{\textbf{Toolchain}} & \color{white}{\textbf{Optimization}} & \color{white}{\textbf{Architecture}}& \color{white}{\textbf{\#Loops}} & \color{white}{\textbf{\#op.}} & \color{white}{\textbf{II}} & \color{white}{\textbf{\#unused PE}} & \color{white}{\textbf{max(\#op. per PE)}} \\\hline\hline
 \rowcolor{darkGray} \multicolumn{8}{|c|}{\color{white}{\textbf{GEMM}}} \\\hline
 {CGRA-Flow}                                          & -                                    & classical CGRA                            & 3                              & 23                            & 10                         & 6                                   & 4 \\ 
 {CGRA-Flow}                                          & flat                                 & classical CGRA                            & 3                              & 28                            & 6                          & 5                                   & 4 \\ 
 {CGRA-Flow}                                          & flat+unroll                          & classical CGRA                            & 3                              & 52                            & 6                          & 0                                   & 6 \\
 {Morpher}                                            & {flat}                               & classical CGRA                            & 3                              & 47                            & 9                          & 4                                   & 9 \\ 
 {Morpher}                                            & {flat}                               & HyCUBE                               & 3                              & 47                            & 9                          & 5                                   & 9 \\ 
 {Morpher}                                            & {flat+unroll}                        & classical CGRA                            & 3                              & 80                            & 8                          & 2                                   & 8 \\ 
 {Morpher}                                            & {flat+unroll}                        & HyCUBE                               & 3                              & 80                            & 8                          & 1                                   & 8 \\
 \rowcolor{darkOrange!50}{CGRA-ME}                    & {-}                                  & HyCUBE                               & 1                              & 23                            & 1                          & 5                                   & 1 \\
 \rowcolor{darkOrange!50}{Pillars}                    & {-}                                  & ADRES                                & 1                              & 23                            & 1                          & 5                                   & 1 \\
 {TURTLE}                                               & {-}                                  & TCPA                               & 3                              & 11                            & 1                          & 0                                   & 11 \\ \hline\hline
 \rowcolor{darkGray} \multicolumn{8}{|c|}{\color{white}{\textbf{ATAX}}} \\\hline
 {CGRA-Flow}                                          & -                                    & classical CGRA                            & 2                              & 30                            & 13                         & 3                                   & 6 \\ 
 {CGRA-Flow}                                          & flat                                 & classical CGRA                            & 2                              & 36                            & 10                         & 0                                   & 4 \\
 {CGRA-Flow}                                          & flat+unroll                          & classical CGRA                            & 2                              & 87                            & 25                         & 0                                   & 8 \\ 
 {Morpher}                                            & {flat}                               & classical CGRA                            & 2                              & 55                            & 14                         & 3                                   & 10 \\
 {Morpher}                                            & {flat}                               & HyCUBE                               & 2                              & 55                            & 10                         & 1                                   & 8 \\
 \rowcolor{darkRed!50}{Morpher}                       & {flat+unroll}                        & classical CGRA                            & 2                              & 118                           & -                          & -                                   & - \\ 
 {Morpher}                                            & {flat+unroll}                        & HyCUBE                               & 2                              & 118                           & 14                         & 0                                   & 14 \\
 \rowcolor{darkOrange!50}{CGRA-ME}                    & {-}                                  & HyCUBE                               & 1                              & 21                            & 2                          & 11                                  & 2 \\
 \rowcolor{darkRed!50}{Pillars}                       & {-}                                  & ADRES                                & 1                              & 21                            & -                          & -                                   & - \\
 {TURTLE}                                               & {-}                                  & TCPA                               & 2                              & 12                            & 3                          & 0                                   & 12 \\ \hline\hline
 \rowcolor{darkGray} \multicolumn{8}{|c|}{\color{white}{\textbf{GESUMMV}}} \\\hline
 {CGRA-Flow}                                          & -                                    & classical CGRA                            & 2                              & 22                            & 8                          & 6                                   & 4\\ 
 {CGRA-Flow}                                          & flat                                 & classical CGRA                            & 2                              & 25                            & 5                          & 3                                   & 4 \\ 
 {CGRA-Flow}                                          & flat+unroll                          & classical CGRA                            & 2                              & 58                            & 7                          & 0                                   & 6 \\
 {Morpher}                                            & {flat}                               & classical CGRA                            & 2                              & 41                            & 6                          & 4                                   & 6 \\ 
 {Morpher}                                            & {flat}                               & HyCUBE                               & 2                              & 41                            & 6                          & 3                                   & 6 \\
 \rowcolor{darkRed!50}{Morpher}                       & {flat+unroll}                        & classical CGRA                            & 2                              & 86                            & -                          & -                                   & - \\ 
 {Morpher}                                            & {flat+unroll}                        & HyCUBE                               & 2                              & 86                            & 9                          & 2                                   & 6 \\ 
 \rowcolor{darkOrange!50}{CGRA-ME}                    & {-}                                  & HyCUBE                               & 1                              & 28                            & 7                          & 10                                  & 3 \\
 \rowcolor{darkRed!50}{Pillars}                       & {-}                                  & ADRES                                & 1                              & 28                            & -                          & -                                   & - \\
 {TURTLE}                                               & {-}                                  & TCPA                               & 2                              & 12                            & 3                          & 0                                   & 12 \\ \hline\hline
 \rowcolor{darkGray} \multicolumn{8}{|c|}{\color{white}{\textbf{MVT}}} \\\hline
 {CGRA-Flow}                                          & -                                    & classical CGRA                            & 2                              & 29                            & 8                          & 3                                   & 5 \\ 
 {CGRA-Flow}                                          & flat                                 & classical CGRA                            & 2                              & 31                            & 5                          & 2                                   & 4 \\
 {CGRA-Flow}                                          & flat+unroll                          & classical CGRA                            & 2                              & 73                            & 7                          & 0                                   & 6 \\
 {Morpher}                                            & {flat}                               & classical CGRA                            & 2                              & 49                            & 7                          & 3                                   & 7 \\ 
 {Morpher}                                            & {flat}                               & HyCUBE                               & 2                              & 49                            & 7                          & 3                                   & 7\\ 
 {Morpher}                                            & {flat+unroll}                        & classical CGRA                            & 2                              & 106                           & 9                          & 0                                   & 9 \\ 
 {Morpher}                                            & {flat+unroll}                        & HyCUBE                               & 2                              & 106                           & 8                          & 0                                   & 8 \\ 
 \rowcolor{darkOrange!50}{CGRA-ME}                    & {-}                                  & HyCUBE                               & 1                              & 21                            & 2                          & 11                                  & 2 \\
 \rowcolor{darkRed!50}{Pillars}                       & {-}                                  & ADRES                                & 1                              & 21                            & -                          & -                                   & - \\
 {TURTLE}                                               & {-}                                  & TCPA                               & 2                              & 13                            & 3                          & 0                                   & 12 \\ \hline\hline
 \rowcolor{darkGray} \multicolumn{8}{|c|}{\color{white}{\textbf{TRISOLV}}} \\\hline
 {CGRA-Flow}                                          & -                                    & classical CGRA                            & 3                              & 27                            & 10                         & 3                                   & 4\\ 
 \rowcolor{darkRed!50}{CGRA-Flow}                     & flat                                 & classical CGRA                            & 3                              & 44                            & -                          & -                                   & -\\ 
 \rowcolor{darkRed!50}{CGRA-Flow}                     & flat+unroll                          & classical CGRA                            & 3                              & 138                           & -                          & -                                   & -\\ 
 {Morpher}                                            & {flat}                               & classical CGRA                            & 3                              & 57                            & 8                          & 3                                   & 7 \\ 
 {Morpher}                                            & {flat}                               & HyCUBE                               & 3                              & 57                            & 7                          & 3                                   & 4 \\
 \rowcolor{darkRed!50}{Morpher}                       & {flat+unroll}                        & classical CGRA                            & 3                              & 180                           & -                          & -                                   & -\\
 \rowcolor{darkRed!50}{Morpher}                       & {flat+unroll}                        & HyCUBE                               & 3                              & 180                           & -                          & -                                   & -\\ 
 \rowcolor{darkOrange!50}{CGRA-ME}                    & {-}                                  & HyCUBE                               & 1                              & 21                            & 2                          & 11                                  & 2 \\
 \rowcolor{darkRed!50}{Pillars}                       & {-}                                  & ADRES                                & 1                              & 21                            & -                          & -                                   & - \\ 
 {TURTLE}                                               & {-}                                  & TCPA                               & 3                              & 11                            & 6                          & 0                                   & 11 \\ \hline
        \end{tabular}
    }
\end{table}

        For performance evaluation, we picked five common loop benchmarks from the Polybench suite~\cite{polybench}.
        Each benchmark is a multidimensional loop nest that represents a typical workload in the domain of linear algebra.
        In the following, we briefly describe each benchmark mathematically assuming that $\mmat{A}, \mmat{B}, \mmat{C}, \mmat{D}$ are ${N\times N}$ matrices and $\mvec{x}, \mvec{x}_1, \mvec{x}_2, \mvec{y}, \mvec{y}_1, \mvec{y}_2$ are vectors of size $N$ with $x_i \in \mvec{x}, y_i \in \mvec{y}$ and $a_{i, j} \in \mmat{A}$.
        \begin{itemize}
            \item GEMM: $\mmat{D} = \mmat{A} \cdot \mmat{B} + \mmat{C}$
            \item ATAX: $\mvec{y} = \mmat{A}\mtrans \cdot (\mmat{A} \cdot \mvec{x})$
            \item GESUMMV: $\mvec{y} = \mmat{A} \cdot \mvec{x} + \mmat{B} \cdot \mvec{x}$
            \item MVT: $\mvec{z}_1 = \mvec{x}_1 + \mmat{A} \cdot \mvec{y}_1;\mvec{z}_2 = \mvec{x}_2 + \mmat{A}\mtrans \cdot \mvec{y}_2$
            \item TRISOLV: $x_i = \frac{y_i - \sum_{j = 0}^{i - 1}x_j \cdot a_{j, i} + \sum_{j = i + 1}^{N - 1}x_j \cdot a_{j, i}}{a_{i, i}}$
        \end{itemize}

        First, we observed that several CGRA toolchains do not accept as input a multidimensional loop nest directly, but require flattening, \ie, the multidimensional loop nest is reduced into a single loop by unfolding the iterations of the outer loops.
        Furthermore, no considered CGRA toolchain unrolls a given loop automatically; thus, this transformation was done manually.
        Then, we mapped each benchmark kernel onto a $4 \times 4$ processor array using the four selected CGRA toolchains, CGRA-Flow~\cite{4_OpenCGRA}, Morpher~\cite{6_MorpherWOSET}, CGRA-ME~\cite{RaghebWWBRYA24}, and Pillars~\cite{guo-pillars-woset2020}, together with TURTLE for TCPAs~\cite{WitteraufWHT21}.
        \tableRef{table:tcpa_cgra_mapping} summarizes the mapping results.
        Whenever no mapping could be found, the entry is colored red, while an orange row indicates a successful mapping of only the innermost loop.

        CGRA-Flow is the only CGRA tool that directly supports multidimensional loops.
        However, we can observe that flattening should still be applied in all but one benchmark as it favorably reduces the achievable initiation interval.
        Only in the case of TRISOLV, the flattened benchmark could not be mapped by CGRA-Flow.
        In this evaluation, we used two different target architectures for Morpher.
        A classical CGRA without multi-hop connections and HyCube, for which Morpher finds consistently better mappings.
        We targeted an ADRES-like~\cite{MeiVVML03} architecture for Pillars, and since Pillars does not come with its own DFG generator, we utilized the DFG from CGRA-ME.
        However, it could only find a valid mapping for the GEMM kernel, failing for all others.
        Without any direct support for multidimensional loops, the loop must be flattened into a single loop.
        This requires explicitly inserting conditional statements inside the loop body that update the outer loop indices.
        Moreover, CGRA-ME currently does not support any predication; hence, it only maps the innermost loop.
        But this simplification allows this tool to achieve the lowest initiation interval \II among the CGRA toolchains.
        However, as discussed in \secRef{sec:cgra:mapping}, the generation of the loop indices should introduce a RecMII of 3.
        This does not apply to CGRA-ME because it only maps the innermost loop and omits any loop-bound checks.
        Besides the achieved initiation interval, the table also shows the number of PEs that were not utilized and the maximum number of operations mapped to a PE.
        It was rarely possible to employ all the 16 available PEs.
        Moreover, the number of operations per PE indicates that even the used PEs are underutilized.
        For example, with $\II=10$ and a maximum of 4 operations per PE, the most active PE will only execute 4 operations within a window of 10 cycles.
        TURTLE can map multidimensional loop nests directly on all available PEs and consistently achieves the best $\II$ among all toolchains, excluding CGRA-ME and Pillars due to the reasons described above.
    \input{figures/performance_figs.tex}
        The achieved $\II$ reflects the mapping quality, but to assess the actual performance, we also investigate for each benchmark the latencies for different input matrix sizes for the best CGRA mappings achieved by Morpher and CGRA-Flow together with the mapping found for the TCPA.
        The achievable latencies are shown in~\figRef{fig:benchmark}, whereas the latencies of the first and last PE in the TCPA to complete are shown separately, because a TCPA could allow to start a next call of the accelerator already after the latency of the first PE and not wait until the last PE is completed.
        Additionally, \figRef{fig:barChart} shows the latency of CGRA schedules for each benchmark normalized to a TURTLE-generated loop nest schedule for a TCPA target of equal size $4\times 4$ PEs.
        For the GEMM kernel, we observe that CGRA-Flow beats Morpher but is outclassed by the TCPA that is \factor{19} faster.
        This is the result of each of the 16 PEs starting a new iteration every single cycle, while in the CGRA every 6 cycles only 4 unrolled iterations are started.
        Similarly, the TCPA outperforms the CGRAs for ATAX and GESUMMV by huge factors of \factor{4} and \factor{3}, respectively, and by \factor{2} for both MVT and TRISOLV.
        Here, we see a gap between the latency of the first and last PE.
        This is due to the fact that both MVT and TRISOLV are only two-dimensional algorithms that, mapped on a two-dimensional array of PEs, cannot be executed entirely in parallel.
        Thus, all PEs are used, but the first PEs already terminate much earlier.
        To investigate this futher, an additional experiment with a TRSM kernel, a three-dimensional loop that executes TRISOLV in the two innermost loops, was performed.
        This utilizes the PEs better, \ie, \factor{8} faster than the next best CGRA mapping found by Morpher, as indicated by the nearly identical latency of the first and last PE.
        However, considering the fact that an application might invoke the same kernel execution multiple times in a row, as shown in~\cite{blu_paper}, the latency to complete one invocation is not as important as the earliest time at which the next invocation can be started.
        This time is the shown latency of the first PE, thus, whenever multiple independent invocations of the same loop program are considered, the TCPA outperforms the CGRAs even more.
        Such overlapped execution is not unique to TCPAs, but was not available on the considered CGRAs.
        In summary, the performance evaluation shows that a TCPA outperforms a CGRA with the same number of PEs.
        However, the PE architecture of CGRAs and TCPAs differ significantly.
        This necessitates an analysis of the hardware costs to put the achieved performance into perspective.

    \subsection{Area}
    \label{sec:quant:area}
        In the following, we also determine and compare the area cost of both architectures for equal size of the processor array (number of PEs).
        For this purpose, we synthesize a generic $4 \times 4$ PE CGRA as well as a $4 \times 4$ TCPA to an FPGA.
        Later, we also compare area margins of actual chip designs.
        \begin{table}
    \caption{%
        Resource utilization of a generic $4\times 4$ CGRA and a $4\times 4$ TCPA.
    }
    \label{table:tcpa_cgra_fpga_area}
    \centering
    \resizebox{0.5\textwidth}{!}{
        \definecolor{darkGray}{RGB}{71, 79, 82}   
        \rowcolors{1}{darkGray!20}{darkGray!5}
        \begin{tabular}{|l|c|c|c|c|c|}
            \rowcolor{darkGray}
            \color{white}{\small \textbf{}} &
            \color{white}{\small \textbf{Insts.}} &
            \color{white}{\small \textbf{LUTs}} &
            \color{white}{\small \textbf{FFs}} &
            \color{white}{\small \textbf{BRAMs}} &
            \color{white}{\small \textbf{DSPs}} \\\hline\hline
            
            \rowcolor{darkGray} \multicolumn{6}{|c|}{\color{white}{\textbf{CGRA}}} \\\hline
            $4\times 4$ CGRA                                                                & \num{1} & \num{35250}  & \num{32552}  & \num{20}  & \num{48} \\
            Avg. Processing element~(PE)                                                    & \num{16} & \num{2202}   & \num{2034}   & \num{1}   & \num{3}  \\
            Avg. ALU~(without division)                                                     & \num{16} & \num{505}    & \num{102}    & \num{0}   & \num{3}  \\
            Avg. Divider                                                                    & \num{16} & \num{1293}   & \num{1629}   & \num{0}   & \num{0}  \\
            Avg. Instruction memory and decoder                                             & \num{16} & \num{400}    & \num{16}     & \num{1}   & \num{0}  \\
            Scratchpad memory (multi bank)                                                  & \num{1} & \num{37}     & \num{2}      & \num{4}   & \num{0}  \\\hline\hline
            \rowcolor{darkGray} \multicolumn{6}{|c|}{\color{white}{\textbf{TCPA}}} \\\hline
            $4\times 4$ TCPA                                                                & \num{1} & \num{220524} & \num{205774} & \num{656} & \num{48} \\
            Avg. Processing element~(PE)                                                    & \num{16} & \num{11091}  & \num{8563}   & \num{39}  & \num{3}  \\ 
            Avg. Functional units                                                           & \num{16} & \num{2967}   & \num{3380}   & \num{7}   & \num{3}  \\
            Avg. Data register file                                                         & \num{16} & \num{6000}   & \num{2947}   & \num{2}   & \num{0}  \\
            Avg. Control register file                                                      & \num{16} & \num{645}   & \num{711}    & \num{30}  & \num{0}  \\
            Avg. Interconnect                                                               & \num{16} & \num{712}   & \num{683}     & \num{0}   & \num{0}  \\ 
            Avg. I/O buffer incl. AGs                                                       & \num{4} & \num{6523}   & \num{11197}  & \num{8}   & \num{0}  \\
            Avg. Address Generator                                                          & \num{32} & \num{483}    & \num{740}    & \num{0}   & \num{0}  \\ 
            Global controller                                                               & \num{1} & \num{9741}   & \num{17861}  & \num{0}   & \num{0}  \\
            Loop I/O controller                                                             & \num{1} & \num{5738}   & \num{4277}   & \num{4}   & \num{0}  \\\hline
        \end{tabular}
    }
\end{table}

        \subsubsection{FPGA Resource Requirements}
            We obtain the resource utilization directly by synthesizing an RTL description for an AMD/Xilinx Ultrascale+ FPGA target using Vivado.
            Most toolchains already provide synthesizable RTL, but for a fair comparison we used in this work for the TCPA the RTL provided by TURTLE and developed for the CGRAs a generic architecture in VHDL as shown in \figRef{fig:mainFigCgra} (right).
            This generic CGRA implements the bare minimum architecture required to execute the loop mappings of the previously evaluated benchmarks in \secRef{sec:quant:performance}.
            Moreover, it resembles the HyCube architecture~\cite{2_HyCUBE}, \ie, it features a single-cycle ALU per PE, a single input and output channel to each neighbor PE, no local register file but 10 multiplexed registers along the data path, and an instruction memory containing up to 16 cycle-by-cycle configurations.
            This configuration is necessary to map all of the previously introduced benchmarks.
            As shown in \figRef{fig:mainFigCgra}, only the leftmost PEs have access to a scratchpad memory that uses a multi bank approach where each left border PE has its own distinct \qty{4}{\kilo\byte} memory bank.
            The single ALU in each PE supports support addition, multiplication, and division on \qty{32}{\bit} integer.
            Besides these arithmetic operations, the CGRA also requires each PE to perform basic logic~(and, or, etc.), comparison and load/store operations.
            All operations are implemented as single-cycle operations except the division which takes 16 cycles.
            Similarly, we chose the TCPA parameters such that it is able to execute the previously discussed benchmarks accordingly.
            As a result, each PE contains the following functional units: two adder, one multiplier, one divider, and 3 copy units for moving data between registers.
            Each unit possesses its own instruction pipeline and runs a separate program.
            These programs are stored in FU-local instruction memories of the following size.
            \memSize{78}{47} and \memSize{25}{43} for the adders, \memSize{51}{45} for the multiplier, \memSize{29}{43} for the divider, and \memSize{20}{43} for each copy unit.
            The actual arithmetic units, \ie, adder, multiplier, and divider, are the same as in the generic CGRA.
            Furthermore, the data register file in each PE is chosen to contain in total 32 addressable registers~(8 general-purpose, 8 feedback, 8 input, and 8 output registers) that can be written by all FUs simultaneously.
            Note that the feedback and input registers are FIFOs, with a combined total capacity of \memSize{280}{32} per PE.
            The array is surrounded by the I/O buffers containing a total of 32 $\qty{512}{\byte}$ banks and 32 AGs that are configured, similar to the other peripherals~(GC, LION), to control the schedule of up to 4 loop dimensions.
            Finally, each PE is configured to have 8 channels to each of its neighbors.
            With these assumptions, we synthesized and compared CGRA and TCPA architectures achieving similar clock frequencies of \qtyrange{200}{250}{\mega\hertz}.

            \tableRef{table:tcpa_cgra_fpga_area} shows the resource requirements of the previously discussed generic CGRA architecture implementing a $4 \times 4$ array of PEs.
            As can be observed, the costs of the scratchpad memory next to the array are negligible.
            Furthermore, the average cost of each PE and its main components are listed.
            \qty{58}{\percent} of LUTs and \qty{80}{\percent} of FFs are used for the divider, while the remaining resources are equally distributed among the ALU and the instruction pipeline including memory and decoder.
            \tableRef{table:tcpa_cgra_fpga_area} also shows the costs of a $4 \times 4$ TCPA and its components.
            We observe that \qty{80.47}{\percent} of LUTs and \qty{66.58}{\percent} of FFs are used within the PE array making each TCPA PE approximately 5 times more costly than that of the generic CGRA.
            Most resources within a PE are needed to implement the data register file~(\qty{6000}{\luts}, \qty{2947}{\ffs}) to keep intermediate local data created in iterations and reused in each PE at later iterations due to tile-based assignment of iteration spaces to PEs (iteration-centric mapping).
            The other dominant area requirements come from the FUs~(\qty{2967}{\luts}, \qty{3380}{\ffs}).
            These costs arise from the implementation of virtual registers that enable all 7 FUs to simultaneously write to any register within the data register file.
            For the control generation, each PE requires also a control register file which internal registers could be mapped to BRAM.
            Additionally, the GC that feeds these control register files with control signals comes at a cost of \qty{9741}{\luts} and \qty{17861}{\ffs}, but is required only once for the entire processor array.
            Finally, the I/O buffers at each border are rather cheap~(\qty{6523}{\luts}, \qty{11197}{\ffs}) containing 8 AGs each requiring \qty{483}{\luts} and \qty{740}{\ffs}.
            Finally, the LION used to fill and drain the I/O buffers costs about half of a PE, \ie, \qty{5738}{\luts} and \qty{4277}{\ffs}.
            In summary, this $4 \times 4$ TCPA architecture requires \factor{6.26} the resources of the simple generic $4\times 4$ CGRA architecture.
        \subsubsection{ASIC Area}
            For TCPAs, \cite{alpaca} presents an actual $8 \times 8$, \qty{10}{\milli\meter\squared} chip manufactured in \qty{22}{\nano\meter}.
            Meanwhile, there exists also work on CGRAs with \cite{WangKMMP19} presenting a CGRA with 16 PEs on \qty{4.7}{\milli\meter\squared} in \qty{40}{\nano\meter} and \cite{FengCKKLMNSZNST22} showing a \qty{16}{\nano\meter} chip containing 384 PEs within an area of \qty{20.1}{\milli\meter\squared}.
            We may compare these chips by normalizing the chip area to the PE count and technology size.
            We use a scaling factor of $1.89$ and $6.25$ for \qty{22}{\nano\meter} and \qty{40}{\nano\meter}, respectively.
            This results in a normalized area per PE of \qty{0.083}{\milli\meter\squared} for \cite{alpaca}~(TCPA), \qty{0.047}{\milli\meter\squared} for \cite{WangKMMP19}~(CGRA), and \qty{0.052}{\milli\meter\squared} for \cite{FengCKKLMNSZNST22}~(CGRA).
            Note that the supported number format of the FUs inside each PE is different in each of those chips.
            In particular, the FUs of the TCPA architecture in \cite{alpaca} support \qty{32}{\bit} floating point, while the CGRAs in \cite{WangKMMP19, FengCKKLMNSZNST22} only support \qty{32}{\bit} fixed point and  both \qty{16}{\bit} bfloat and \qty{16}{\bit} integer, respectively.
    \subsection{Power}
    \label{sec:quant:power}
        Analogously to the area analysis in \secRef{sec:quant:area}, we investigate for both CGRAs and TCPA the power consumption on FPGAs and summarize results from literature for ASICs.
        \subsubsection{FPGA Power}
            The vectorless power analyzer of Vivado reported for the two $4 \times 4$ architectures presented above a power consumption of \qty{3.313}{\watt} for the TCPA and \qty{1.957}{\watt} for the CGRA.
            Surprisingly, the TCPA design requiring \factor{6.26} the resources only consumes \factor{1.69} the power.
        \subsubsection{ASIC Power}
            The authors of the previously discussed published chip designs for CGRAs and TCPAs also report power and energy efficiency.
            The $8\times 8$ floating-point TCPA in~\cite{alpaca} consumes \qty{7.5}{\watt} at peak~(\qty{117}{\milli\watt} per PE), while the CGRA in~\cite{WangKMMP19} has a peak power consumption of \qty{102}{\milli\watt} in total and \qty{6.375}{\milli\watt} per PE.
            \cite{FengCKKLMNSZNST22} only reports a peak energy efficiency of \qty[per-mode=repeated-symbol]{538}{\giga\ops\per\watt}, but not the actual power.
            \cite{WangKMMP19}~(CGRA) and \cite{alpaca}~(TCPA) show a peak energy efficiency of \qty[per-mode=repeated-symbol]{26.4}{\giga\ops\per\watt} and \qty[per-mode=repeated-symbol]{270}{\giga\flops\per\watt}, respectively.
\section{Discussion}
\label{sec:discussion}
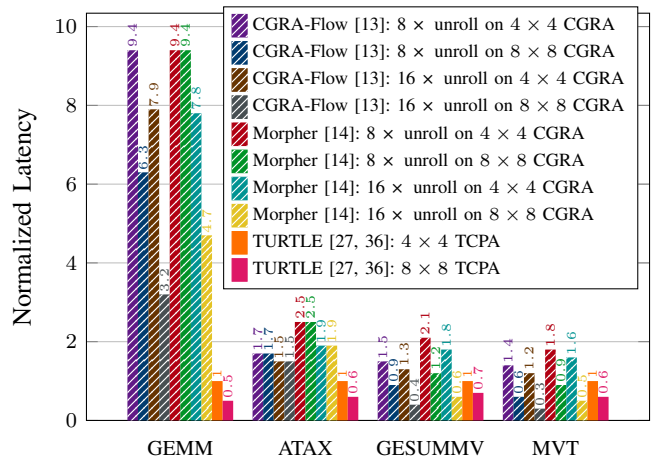
\begin{figure}
    \begin{tikzpicture}
        \definecolor{darkPurple}{RGB}{96, 25, 134}
        \definecolor{darkBlue}{RGB}{0, 59, 111}
        \definecolor{darkBrown}{RGB}{102, 51, 0}
        \definecolor{darkGray}{RGB}{71, 79, 82}
        \definecolor{darkRed}{RGB}{177, 0, 18}
        \definecolor{darkGreen}{RGB}{0, 147, 39}
        \definecolor{darkCyan}{RGB}{0, 147, 146}
        \definecolor{darkOrange}{RGB}{255, 111, 0}
        \definecolor{darkYellow}{RGB}{227, 195, 36}
        \definecolor{darkRosa}{RGB}{222, 22, 102}
        \begin{axis}[
            width  = 0.5*\textwidth,
            height = 7cm,
            major x tick style = transparent,
            ybar=0pt, 
            bar width=4pt,
            ymajorgrids = true,
            ylabel = {Normalized Latency},
            tick label style = {font=\footnotesize},
            symbolic x coords={GEMM, ATAX, GESUMMV, MVT},
            xtick = data,
            scaled y ticks = false,
            enlarge x limits=0.25,
            ymin=0,
            legend image code/.code={
            \draw [#1] (0cm,-0.15cm) rectangle (0.2cm,0.15cm); },
            legend style={font=\scriptsize,yshift=0.2cm},
            legend cell align={left},
            nodes near coords,
            nodes near coords style = {font=\tiny,rotate=90,anchor=west,inner sep=0.03cm},
        ]
            \addplot +[style={darkPurple, draw=none}, postaction={pattern=north east lines, pattern color=white}]
                coordinates {(GEMM,9.4) (ATAX,1.7) (GESUMMV,1.5) (MVT,1.4)};
            \addplot +[style={darkBlue, draw=none}, postaction={pattern=north east lines, pattern color=white}]
                coordinates {(GEMM,6.3) (ATAX,1.7) (GESUMMV,0.9) (MVT,0.6)};
            \addplot +[style={darkBrown, draw=none}, postaction={pattern=north east lines, pattern color=white}]
                coordinates {(GEMM,7.9) (ATAX,1.5) (GESUMMV,1.3) (MVT,1.2)};
            \addplot +[style={darkGray, draw=none}, postaction={pattern=north east lines, pattern color=white}]
                coordinates {(GEMM,3.2) (ATAX,1.5) (GESUMMV,0.4) (MVT,0.3)};
            \addplot +[style={darkRed, draw=none}, postaction={pattern=north east lines, pattern color=white}]
                coordinates {(GEMM,9.4) (ATAX,2.5) (GESUMMV,2.1) (MVT,1.8)};
            \addplot +[style={darkGreen, draw=none}, postaction={pattern=north east lines, pattern color=white}]
                coordinates {(GEMM,9.4) (ATAX,2.5) (GESUMMV,1.2) (MVT,0.9)};
            \addplot +[style={darkCyan, draw=none}, postaction={pattern=north east lines, pattern color=white}]
                coordinates {(GEMM,7.8) (ATAX,1.9) (GESUMMV,1.8) (MVT,1.6)};
            \addplot +[style={darkYellow, draw=none}, postaction={pattern=north east lines, pattern color=white}]
                coordinates {(GEMM,4.7) (ATAX,1.9) (GESUMMV,0.6) (MVT,0.5)};
            \addplot +[style={darkOrange, draw=none}]
                coordinates {(GEMM,1) (ATAX,1) (GESUMMV,1) (MVT,1)};
            \addplot +[style={darkRosa, draw=none}]
                coordinates {(GEMM,0.5) (ATAX,0.6) (GESUMMV,0.7) (MVT,0.6)};

            \legend{%
                {CGRA-Flow~\cite{4_OpenCGRA}: \factor{8} unroll on $4\times 4$ CGRA},%
                {CGRA-Flow~\cite{4_OpenCGRA}: \factor{8} unroll on $8\times 8$ CGRA},%
                {CGRA-Flow~\cite{4_OpenCGRA}: \factor{16} unroll on $4\times 4$ CGRA},%
                {CGRA-Flow~\cite{4_OpenCGRA}: \factor{16} unroll on $8\times 8$ CGRA},%
                {Morpher~\cite{6_MorpherWOSET}: \factor{8} unroll on $4\times 4$ CGRA},%
                {Morpher~\cite{6_MorpherWOSET}: \factor{8} unroll on $8\times 8$ CGRA},%
                {Morpher~\cite{6_MorpherWOSET}: \factor{16} unroll on $4\times 4$ CGRA},%
                {Morpher~\cite{6_MorpherWOSET}: \factor{16} unroll on $8\times 8$ CGRA},%
                {TURTLE~\cite{WitteraufWHT21, TanaseHannig2018ATECS}: $4\times 4$ TCPA},%
                {TURTLE~\cite{WitteraufWHT21, TanaseHannig2018ATECS}: $8\times 8$ TCPA}%
            }
        \end{axis}
    \end{tikzpicture}
    \caption{%
        Speedup of TURTLE-compiled loop nests compared to CGRA-Flow~\cite{4_OpenCGRA} and Morpher~\cite{6_MorpherWOSET} for an input matrix size of $20 \times 20$ for GEMM and $32 \times 32$ for all other benchmarks assuming different unroll levels and number of PEs.
        Note that settings, where the respective tool was not able to provide any feasible schedule, are marked in stripe color showing only a theoretical lower bound.
    }
    \label{fig:theoBarChart}
\end{figure}
    The evaluation in~\secRef{sec:quant} assumed the same number of PEs for both CCRA and TCPA.
    This raises the question of scaling up a CGRA in terms of the number of PEs to achieve the same area cost as a TCPA and compare the achievable performance.
    A major advantage of processor arrays is that their area, power, and theoretical performance scale linearly with the number of PEs, since the architecture of a PE is the same for larger arrays and the cost of additional peripheral controllers is typically very small, as shown in \secRef{sec:quant:area}.
    We conducted an additional experiment to assess the performance gain from larger arrays, shown in \figRef{fig:theoBarChart}.
    Similar to~\figRef{fig:barChart}, it shows the speedup between TCPA and CGRA toolchains for different benchmarks.
    In this case, however, the number of PEs and the unroll factor are varied.
    However, since no CGRA toolchain could find an actual mapping for the configurations shown, the figure only shows a theoretical lower-bound latency computed according to the ResMII and the RecMII after DFG generation.
    Therefore, it only shows the best latency the mapping tool could have achieved, but no actual mapping could be determined.
    In contrast, we observe that TURTLE found a mapping for both $4 \times 4$ and $8 \times 8$ arrays.
    A larger array only results in smaller tiles during partitioning, while the mapping complexity itself does not increase.
    Note also that larger TCPAs naturally support larger problems, since the previously discussed problem size limit depends only on the tile size, \ie, the number of iterations per PE.
    However, the performance gain from 16 to 64 PEs is not \factor{4}.
    This is due to the fact that the wave-like starting and stopping of a larger array combined with smaller tiles leads to a worse overall PE utilization, \ie, the difference between the latency of the first and last PE to complete increases.
    Note, however, that the minimum time at which the next problem can be started on the same array is reduced by \factor{4} on the larger array.
    Hence, TCPAs benefit from larger arrays but require either an increase in problem size and, thus, tile size or a batch processing use case where the same kernel is repeatedly executed many times.
    For CGRAs, however, increasing the number of PEs is less effective.
    We observed that without increasing the unroll factor, more PEs only mitigate the ResMII, but do not reduce the RecMII.
    Thus, in many settings, the latency difference between a $4 \times 4$ array and an $8 \times 8$ array, \ie, \factor{4} more PEs, is often zero.
    Even in settings where the difference is significant, only a speedup of \factorRange{2}{3} should be theoretically possible, while \factor{4} hardware resources are required.
    A significant performance gain is only possible by increasing the number of nodes within the DFG by further unrolling the loop.
    Here, we observe that still no evaluated CGRA approach is able to find lower latency mappings than found for the TCPA for the GEMM and ATAX benchmarks.
    Only for the GESUMMV and MVT benchmarks, an $8\times 8$ CGRA can theoretically outperform a $4 \times 4$ TCPA if the unrolling factor is also chosen sufficiently large.
    However, the DFGs in these cases become large with up to 330 nodes, which makes the mapping a massively complex challenge that none of the analyzed tools could handle.
    In theory, it should be possible to map a larger DFG onto a larger array, but this assumes that the nodes can be evenly distributed across the available resources.
    However, this may not be possible due to increased routing congestion around the border PEs.
    In the classic CGRA architecture, only the left border PEs have access to memory.
    Thus, in a $4 \times 4$ array, only $4 / 16$ PEs can issue load and store operations.
    In larger arrays, this factor decreases further, \eg, $8 / 64$ in an $8 \times 8$ array.
    Since unrolling a loop causes the number of load and store operations to increase linearly, routing contention around the border PE will worsen significantly.
    Additional memory banks around all 4 sides of the array mitigate this problem to some extent, but when the array becomes too large, this will not suffice.
    A better solution is the approach used in TCPAs.
    Instead of reading and writing the inputs and outputs of all unrolled iterations within the DFG in every execution, the data must be kept locally within the array and reused within different unrolled iterations of the same DFG.
    In summary, larger CGRAs may only simplify the mapping without a significant performance gain to justify the increase in hardware cost.
    Moreover, even with the poor scaling properties of CGRAs, multiple parallel CGRAs could have the same area cost as a TCPA, assuming there is enough parallelism at the kernel level.
    Note, however, that in this case the TCPA could also exploit its ability to overlap multiple kernel executions, further outperforming CGRAs and their operation-centric mapping approaches, as shown in \secRef{sec:quant:performance}.
\section{Conclusion}
\label{sec:conclusion}
    In this paper, we analyzed two prominent classes of architectures for accelerating nested loops on processor arrays:
    Coarse-Grained Reconfigurable Arrays (CGRAs) with an operation-centric mapping approach, and Tightly-Coupled Processor Arrays (TCPAs) with an iteration-centric mapping approach.
    While the toolchains for CGRAs map operations from a DFG to PEs, a TCPA mapper tiles a given $n$-dimensional iteration space into as many congruent tiles as available PEs and schedules these iterations both globally and locally to best exploit multiple levels of parallelism and data locality.
    This study provides a comprehensive qualitative and quantitative comparison of four CGRA toolchains and one TCPA toolchain.
    The qualitative evaluation shows that CGRAs may be more intuitive to use, especially for developers familiar with C/C++, but their mapping process often struggles with scalability, requiring significant time for large arrays or complex problems.
    In contrast, compiling loop nests to TCPAs is independent of the size of the iteration space or the size of the target processor array.
    Optimal schedules can even be determined independent of the size of the loop nest bounds, as shown in~\cite{WitteraufTeich2016ASAP, TanaseHannig2018ATECS}.
    The quantitative evaluation showed significant differences in performance, power, and area (PPA) due to the different mapping methods of these architectures.
    In TCPAs, the PEs must be able to execute a full tile of iterations.
    This requires a more complex PE architecture with local memory in the form of a register file, multiple FUs to locally execute all operations within the loop body, and multiple interconnect channels to neighboring PEs.
    Special feedback registers also distinguish a TCPA PE from other PE architectures.
    These support the efficient reuse of data computed in earlier iterations within a tile, without the need to communicate intermediate data to external memories.
    Obviously, this makes the PE architecture used in CGRAs\,---\,basically an ALU, a crossbar, some registers, and an instruction memory with a decoder\,---\,smaller because it only has to execute single operations, not entire loop bodies, and not multiple iterations.
    In addition, TCPAs are designed to offload all control overhead, such as loop counter incrementing, memory access address generation, loop bound tests, etc., to additional control units outside the core array.
    %
    But despite an approximately \factor{5} higher resource count of a TCPA when synthesized on an FPGA, the resulting power consumption is only \factor{1.69} higher than that of a generic CGRA.
    More importantly, this additional cost translates into significant performance gains.
    In our experiments with five common benchmarks, TCPAs consistently outperformed CGRAs, achieving up to a \factor{19} speedup on the GEMM benchmark.
    This is due to several factors:
    In many CGRA mappings, several PEs remain completely inactive, \ie, no operations are mapped to them, while the maximum number of operations mapped to a PE further indicates that these PEs are not well utilized, one reason being a lack of routing opportunities.
    For example, we observe that the HyCube architecture, which increases routing capability by introducing multi-hop connections, consistently outperforms the classic CGRAs without multi-hop connections.
    In addition, the PEs of a CGRA must perform all control flow and address computation, often contributing to more than \qty{70}{\percent} of the operations in common loop programs, as illustrated in \figRef{fig:mainFigCgra}.
    Also, the generated DFGs often contain fatal throughput-limiting cyclic dependencies.
    We believe that in the future, the pure operation-centric approach used in CGRAs will be combined with some iteration-centric methods, \eg, extensions similar to~\cite{KoulMSTNZLSCMSDDCKFHNSTBDMTRBFHBHT23} that separate control flow from data flow.
    Moreover, for TCPAs, automatic single-assignment generation from imperative for loop descriptions, e.g., \cite{Feautrier96}, may make these architectures more attractive from the standpoint of programmability and usability.

\vspace{8pt}
{\small%
\noindent%
\textbf{Acknowledgment:} This work was funded by the Deutsche Forschungsgemeinschaft (DFG, German Research Foundation) -- Project number 146371743 -- TRR 89: Invasive Computing.%
}

\renewcommand*{\bibfont}{\scalefont{0.93}}
\printbibliography

\end{document}